\documentclass[fleqn,usenatbib]{mnras}
\usepackage[english]{babel}

\usepackage[T1]{fontenc}
\usepackage{ae,aecompl}

\usepackage{fancyhdr}
\usepackage{amsfonts}
\usepackage{amsmath}
\usepackage{amssymb}
\usepackage{multicol}
\usepackage{layout}
\usepackage{graphicx}
\usepackage{multirow}

\usepackage{color}
\usepackage{multirow}

\usepackage{times}
\usepackage{natbib}

\newif\ifAMStwofonts
\AMStwofontstrue

\graphicspath{{./fig/}}

\title[Growth of spherical overdensities in ST gravities]{Growth of spherical overdensities in scalar-tensor cosmologies}

\author[Nazari-Pooya et~al.]{N. Nazari-Pooya$^{1}$,
M. Malekjani$^{2}$ \thanks{malekjani@basu.ac.ir}, F. Pace$^{3}$ and D. Mohammad-Zadeh Jassur$^{1}$\\
$^1$ Department of Theoretical Physics and Astrophysics, Tabriz University, Tabriz, Iran.\\
$^2$ Department of Physics, Bu-Ali Sina University, Hamedan 65178, Iran.\\
$^3$ Jodrell Bank Centre for Astrophysics, School of Physics and Astronomy, The University of Manchester, Manchester,
M13 9PL, U.K.}

\date{Accepted ?, Received ?; in original form \today}

\pagerange{\pageref{firstpage}--\pageref{lastpage}} \pubyear{2015}

\begin{document}

\label{firstpage}

\maketitle

\begin{abstract}
The accelerated expansion of the universe is a rather established fact in cosmology and many different models have
been proposed as a viable explanation. Many of these models are based on the standard general relativistic framework
of non-interacting fluids or more recently of coupled (interacting) dark energy models, where dark energy (the scalar
field) is coupled to the dark matter component giving rise to a fifth-force. An interesting alternative is to couple
the scalar field directly to the gravity sector via the Ricci scalar. These models are dubbed non-minimally coupled
models and give rise to a time-dependent gravitational constant. In this work we study few models falling into this
category and describe how observables depend on the strength of the coupling. We extend recent work on the subject
by taking into account also the effects of the perturbations of the scalar field and showing their relative importance
on the evolution of the mass function. By working in the framework of the spherical collapse model, we show that
perturbations of the scalar field have a limited impact on the growth factor (for small coupling constant) and on the
mass function with respect to the case where perturbations are neglected.
\end{abstract}


\section{Introduction}
%Accelerating expansion and EoS parameter
%%%%%%%%%%%%%%%%%%%%%%%%%%%%%%%%%%%%%%%%%%%%%%%%%%%%%%%%%%%%%%%
Geometrical probes of the cosmic expansion scenario including:
(a) Type Ia supernovae (SnIa)~\citep{Riess1998,Perlmutter1999,Kowalski2008},
(b) The first peak location in the angular power spectrum of CMB perturbations \citep{Komatsu2009},
(c) Baryon acoustic oscillations in the power spectrum of the matter density field \citep{Percival2010},
(d) Observations from Gamma Ray Bursts \citep{Basilakos2008b}, cluster gas mass fraction \citep{Allen2004} and
the estimation of the age of Universe \citep{Krauss2003},
and dynamical probes of the growth rate of matter perturbations such as
(e) The growth data from X-Ray clusters \citep{Mantz2008}
(f) The power spectrum at different redshift slices of the Ly-$\alpha$ forest \citep{McDonald2005,Nesseris2008},
(g) Redshift distortions from anisotropic pattern of galactic redshifts at the cluster scales
\citep{Hawkins2003,Nesseris2008} and
(h) Weak lensing surveys \citep{Benjamin2007,Amendola2008a,Fu2008}
point towards the general conclusion that our Universe is experiencing an accelerated expansion.

This cosmic expansion scenario can be well explained by assuming an additional cosmic fluid with a positive energy
density and a sufficiently negative pressure usually dubbed dark energy (DE). The nature of DE is still unknown and it
seems that new physics beyond the standard model of particle physics is required to explain its properties. Einstein
cosmological constant $\Lambda$ with constant equation of state $w_{\Lambda}=P_{\Lambda}/\rho_{\Lambda}=-1$ is the
earliest and simplest candidate for DE. However, this model suffers from severe theoretical problems, as the
fine-tuning and the cosmic coincidence problems
\citep{Weinberg1989,Sahni2000,Carroll2001,Padmanabhan2003,Peebles2003,Copeland2006}.
Alternatively, scalar fields are plausible candidates for dark energy \citep{Ratra1988a,Wetterich1988}. An outstanding
feature of scalar fields is that the equation of state of these models is generally varying during the cosmic history.
Moreover, due to the time varying equation of state, these models possess some fluctuations in both space and time
like pressureless dust matter.
Several attempts have been done to study the possibility that scalar fields might be coupled to other entities in
the Universe. On one side, the coupling can be considered between scalar fields and dust matter within the framework
of General Relativity (GR). On the other side, one can consider the coupling between the scalar field and the Ricci
scalar, the so-called non-minimally coupled quintessence models, which is the case, for example, of scalar tensor (ST)
gravities \citep{Bergmann1968,Nordtvedt1970,Wagoner1970,EspositoFarese2001}.
For an historical review of ST theories see \cite{Brans1961} and for later attempts
\cite{Perrotta2000,Acquaviva2004,Pettorino2008}. Very recent parameter estimation on Brans-Dicke models have been
performed by \cite{Li2015}.

Beside causing the acceleration of the overall expansion rate of the Universe, quintessence models can change the
formation rate and the growth of collapsed structures (haloes). It is well known that the large scale structures we
observe today originated from the small initial fluctuations originated during the inflationary phase era
\citep{Starobinsky1980,Guth1981,Linde1990}. These fluctuations subsequently grew under the influence of gravity
\citep{Gunn1972,Press1974,White1978,Peebles1993,Peacock1999,Sheth1999,Barkana2001,Peebles2003,Ciardi2005,Bromm2011}.

The spherical collapse model (SCM) introduced by \cite{Gunn1972} is a simple analytical tool to study the evolution of
the growth of overdense structures on sub-Horizon scales. The dynamics of the overdensities depends strongly on the
dynamics of the background Hubble flow and expansion history of the Universe. In the framework of GR, the SCM has been
widely investigated in the literature
\citep{Fillmore1984,Bertschinger1985,Hoffman1985,Ryden1987,AvilaReese1998,Subramanian2000,Ascasibar2004,Williams2004}.
Furthermore, this formalism has been extended to different DE and scalar field DE cosmologies
\citep{Mota2004,Maor2005,Basilakos2009,Li2009,Pace2010,Wintergerst2010,Basse2011,Pace2012,Pace2014b,Naderi2015}.
The growth of spherical overdensities within the framework of inhomogeneous DE cosmologies within the GR framework has
been studied by \cite{Abramo2007,Abramo2009a,Pace2014b,Malekjani2015}.

All of the previously mentioned studies and improvements of the SCM in DE cosmologies take place within the framework
of standard Einstein theory of gravity (GR). In particular, \cite{Mota2004} studied the SCM for different minimally
coupled quintessence models with different potentials in the GR paradigm. However, since in scalar-tensor theories
there is a non-minimally coupling between the scalar field and the gravity sector via the Ricci scalar, the evolution
of spherical overdensities is completely different from that of standard gravity. This is the subject of the work by
\cite{Pace2014} and \cite{Fan2015} where the authors studied the non-linear evolution of structures in non-minimally
coupled quintessence models using the SCM machinery in the context of ST gravities. It is worth to mention that in
\cite{Pace2014}, despite the general derivation of the equations of motion, scalar field perturbations were assumed to
be negligible to facilitate the comparison with results from N-body simulations. In the work by \cite{Fan2015}, the
non-minimally coupled quintessence models are also assumed to be homogeneous in both the metric and Palatini
formalisms.

In this work we extend the SCM in ST theory of gravity to the more general case where scalar field perturbations
are taken into account. In fact in the case of GR, it was shown that density perturbations of minimally coupled scalar
fields exist on all scales but they are strongly scale-dependent and negligible on sub-Hubble scales
\citep{Unnikrishnan2008a,Jassal2009,Jassal2010}. On Hubble scale, they are roughly $10\%$ of the matter density
perturbations and they leave a trace on the low $\ell$ multipoles of the angular power spectrum of the CMB through the
ISW effect \citep{Weller2003a}. Generally, the existence of perturbations of a cosmic fluid on sub-Hubble scales
depends on the effective sound speed $c_{\rm eff}$. The effective sound speed $c_{\rm eff}$ determines the sound
horizon of a fluid: $l_{\rm eff}=c_{\rm eff}/H$. On scales smaller than the sound horizon, perturbations can not grow
and vanish, while on scales larger than $l_{\rm eff}$ perturbations can grow due to gravitational instability. In the
case of minimally coupled quintessence models in GR, the effective sound speed is roughly equal to unity (in units of
the speed of light $c=1$). Consequently, the sound horizon is of the order of the Hubble scale $H^{-1}$
\citep{Ferreira1997,Ferreira1998,dePutter2010}
and the perturbations of scalar fields below the horizon vanish and can not grow. In ST gravities, the non-minimally
coupling between scalar fields and curvature perturbations amplifies the scalar field perturbations on sub-Hubble
scales. In addition, it has been shown that in ST theories, scalar field perturbations on sub-horizon scales are
scale-independent and their effective sound speed vanishes \citep{EspositoFarese2001}. As shown by
\cite{BuenoSanchez2010}, scalar field perturbations are anti-correlated to the perturbations of dust matter. In
addition, the ratio of the scalar field density perturbations over the matter density perturbations on sub-Hubble
scales is roughly of the order of $10\%$ \citep{BuenoSanchez2010}.

The aim of this study is to generalise the SCM in ST theories of gravity for clustering non-minimally coupled
quintessence models by taking the scalar field perturbations into account. We follow the evolution of perturbations
both in the linear and non-linear regimes. We calculate the fundamental SCM parameters: the linear overdensity
threshold $\delta_{\rm c}$ and the virial overdensity $\Delta_{\rm vir}$ in the framework of ST gravities for
different homogeneous and clustering cases of non-minimally quintessence models. Having these quantity at hands, we
will use them to evaluate the mass function and the number counts of haloes. We organize the paper as follows: In
section~\ref{sect:ST-theory} we introduce the ST theories of gravity and describe the evolution of the background
cosmologies in these models. In section~\ref{sect:perturbations}, the basic equations for the evolution of the density
perturbations of the scalar field and matter are presented. In section~\ref{sect:growth_structures} we study the linear
growth factor on sub-horizon scales and the SCM in the framework of ST gravities. We also present the effect of scalar
field perturbations on the mass function and cluster number counts within the Press-Schechter formalism. Finally, we
conclude and summarise our results in section~\ref{sect:conclusions}.

%************************************************************************
%***********************************************************************
\section{Background history in ST theories}\label{sect:ST-theory}
In this section we present the background evolution equations of ST gravity in a spatially flat
Friedmann-Robertson-Walker (FRW) universe. The action for these models in the physical Jordan frame is given by
\citep{Bergmann1968,Nordtvedt1970,Wagoner1970}

\begin{equation}
\begin{split}
S=\frac{1}{16\pi G}\int d^4x~\sqrt{-g} \left(F(\Phi)~R -
Z(\Phi)~g^{\mu\nu}
  \partial_{\mu}\Phi\partial_{\nu}\Phi - 2U(\Phi) \right) \\
  + S_{\rm m}(g_{\mu\nu}) \;, \label{eq:S_JF}
\end{split}
\end{equation}
where $G$ is the gravitational coupling constant, $R$ is the Ricci scalar, $g$ is the determinant of the metric
$g_{\mu\nu}$ and $S_{\rm m}$ is the action of the matter field which does not involve the scalar field $\Phi$. The
independence of the matter action $S_{\rm m}$ from the scalar field $\Phi$ guarantees that the weak equivalence
principle is exactly satisfied. $F(\Phi)$ and $Z(\Phi)$ in equation~(\ref{eq:S_JF}) are arbitrary dimensionless
functions and $U(\Phi)$ is the scalar field potential. The dynamics of the real scalar field $\Phi$ depends on the
dimensionless functions $F(\Phi)$ and $Z(\Phi)$ as well as the potential $U(\Phi)$. The term $F(\Phi)R$ represents the
non-minimally coupling between the scalar field $\Phi$ and gravity. In the limit of GR, it is obvious to have
$F(\Phi)=1$, showing that there is no direct interaction between the scalar field and gravity. By a redefinition
of the field $\Phi$, the quantity $Z(\Phi)$ can be set either to $1$ or $-1$. In this work we will consider all
equations and quantities in ST gravity for the case $Z(\Phi)=1$. In what follows we will also use units such that
$8\pi G=1$. For a flat FRW universe with
\begin{equation}
 ds^2 = -dt^2 + a^2(t)\left[{dr^2}+r^2 \left(d\theta^2+\sin^2{\theta}~d\phi^2\right)\right]\;, \label{eq:frw}
\end{equation}
Friedmann equations for the evolution of the background in ST cosmologies are the following
\citep[see also][]{Gannouji2006,Polarski2008,BuenoSanchez2010}

\begin{eqnarray}
3F(\Phi) H^2  & = & \rho_{\rm m}+\frac{1}{2}\dot{\Phi}^2+U(\Phi)-3H\dot{F}(\Phi)=\rho_{\rm tot}\;, \label{eq:stbh1}\\
-2F(\Phi) \dot H & = & \rho_{\rm m}+\dot{\Phi}^2+\ddot{F}(\Phi)-H\dot{F}(\Phi)=\rho_{\rm tot}+p_{\rm tot}\;.
\label{eq:stbh2}
\end{eqnarray}

The Klein-Gordon equation for the evolution of the scalar field and the equation of motion for non-relativistic dust
matter are, respectively, given by
\begin{eqnarray}
\ddot{\Phi}+3H\dot{\Phi} & = & 3\frac{dF}{d\Phi}\left(\dot{H}+2H^2\right)-\frac{dU}{d\Phi} \;. \label{eq:stbphi}\\
\dot{\rho}_{\rm m}+3H\rho_{\rm m} & = & 0 \;, \label{eq:stbmat}
\end{eqnarray}
where $\rho_{\rm m}$ is the energy density of pressure-less dust matter. Here we will assume the Ratra-Peebles form
for the scalar field potential:
\begin{equation}
 U(\Phi)=\frac{M^{4+\alpha}}{\Phi^{\alpha}}\;,
 \label{eq:Rat-Peeb}
\end{equation}
where $M$ is an energy scale and the exponent $\alpha$ is a free positive constant. Constraints from magnitude-redshift
measurements of type Ia Supernovae show that $\alpha<1$ at the $1-\sigma$ level both for minimally and non-minimally
coupled quintessence models \citep{Caresia2004}. We further assume a power-law for the the dimensionless function
$F(\Phi)$:
\begin{equation}
 F(\Phi)=1+\xi(\Phi^2-\Phi_0^2)\;,
 \label{eq:f_phi}
\end{equation}
where the constant $\xi$ indicates the strength of the coupling between the scalar field and Ricci scalar and $\Phi_0$
is the value of the scalar field at the present time \citep[see also][]{Perrotta2000,Perrotta2002}. Solar system tests
put very tight constraints on the coupling parameter $\xi\approx 10^{-2}$,
\citep[see][]{Reasenberg1979,Chiba1999,Uzan1999,Riazuelo2002,Bertotti2003}.

From equations~(\ref{eq:stbh1}) and~(\ref{eq:stbh2}), the energy density and pressure of the scalar field in the
framework of ST gravities (in the case of $Z(\Phi)=1$ assumed here) can be easily obtained as follows:
\begin{eqnarray}
 \rho_{\phi} & = & \frac{1}{2}\dot{\Phi}^{2}+U(\Phi)-3H\dot{F}(\Phi)\;, \label{eq:energy_density}\\
 p_{\phi} & = & \frac{1}{2}\dot{\Phi}^{2}-U(\Phi)+\ddot{F}(\Phi)+2H\dot{F}(\Phi)\;. \label{eq:pressure}
\end{eqnarray}
We now solve numerically the coupled system of equations (\ref{eq:stbh1}, \ref{eq:stbh2}, \ref{eq:stbphi} and
\ref{eq:stbmat}) for the above potentials $U(\Phi)$ and $F(\Phi)$. We fix the initial time at matter-radiation
equality epoch $a_{\rm i}\approx 10^{-4}$ and also set the present time values of the scalar field and energy density
of pressureless dust matter as $\Phi_0=1$ and $\Omega_{\rm m,0}=0.30$, respectively. We also need two initial
conditions to solve the Klein-Gordon equation. Following \cite{BuenoSanchez2010}, we set the initial conditions as
$\Phi(a_{\rm i})=0.12$ and $\dot{\Phi}(a_{\rm i})=10^{-5}$. We have two free parameters, $\xi$ and $\alpha$, in this
analysis. It should be emphasized that the initial values of these parameters must be chosen such that the function
$F(\Phi)$ and the scalar field $\Phi$ evolve from their initial values to reach $F(\Phi)=1$ and $\Phi=\Phi_0=1$ at the
present time. Consequently, the amount of energy density of the scalar field $\Omega_{\rm \Phi}$ should be
close to $0.70$ at the present time. On the basis of different values of the exponent parameter of the scalar field
potential, $\alpha$, and the strength of the non-minimally coupling, $\xi$, we adopt four different cosmological
models in the framework of ST gravities presented in Table~(\ref{tab:model}). The concordance $\Lambda$CDM model is
reproduced by setting $\alpha=0$ and $\xi=0$. Also model (3) represents a minimally coupled quintessence model in
which there is no direct coupling between the scalar field and gravity ($\xi=0$).

\begin{table}
 \centering
 \caption[Equations of state]{Different minimally and non-minimally coupled quintessence models considered in this
 work. The parameter $\alpha$ indicates the exponent of the inverse power-law potential and $\xi$ represents the
 strength of the coupling between the scalar field and gravity.}
 \begin{tabular}{lcc}
  \hline
  \hline
  Model        & $\xi$  & $\alpha$\\
  \hline
  Model (1)    & 0.123  & 0.261 \\
  Model (2)    & 0.088  & 0.679 \\
  Model (3)    & 0.000  & 0.877 \\
  Model (4)    & -0.087 & 0.877 \\
  $\Lambda$CDM & 0.000  & 0.000 \\
  \hline
 \label{tab:model}
 \end{tabular}
 \begin{flushleft}
  \vspace{-0.5cm}
  {\small}
 \end{flushleft}
\end{table}

\begin{figure}
 \centering
 \includegraphics[width=8cm]{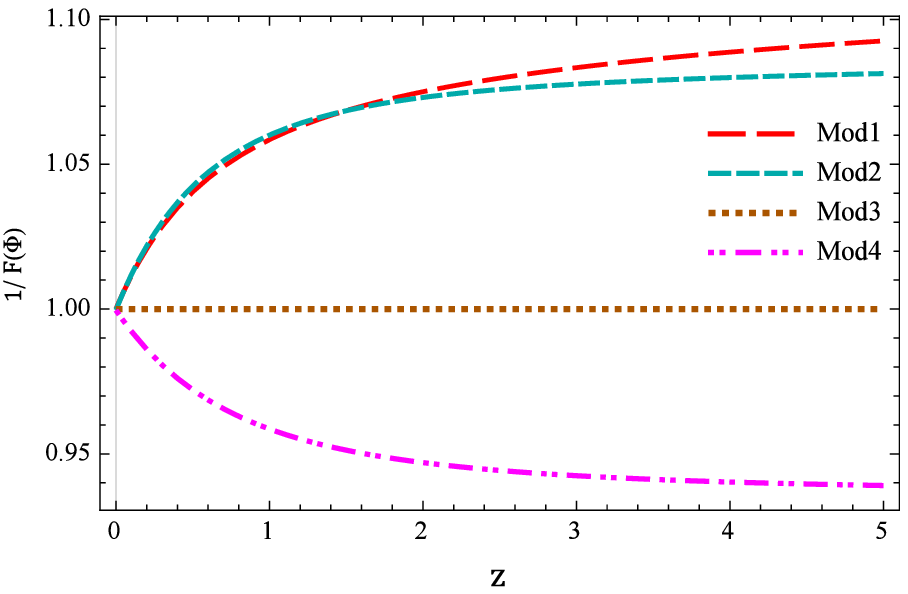}
 \includegraphics[width=8cm]{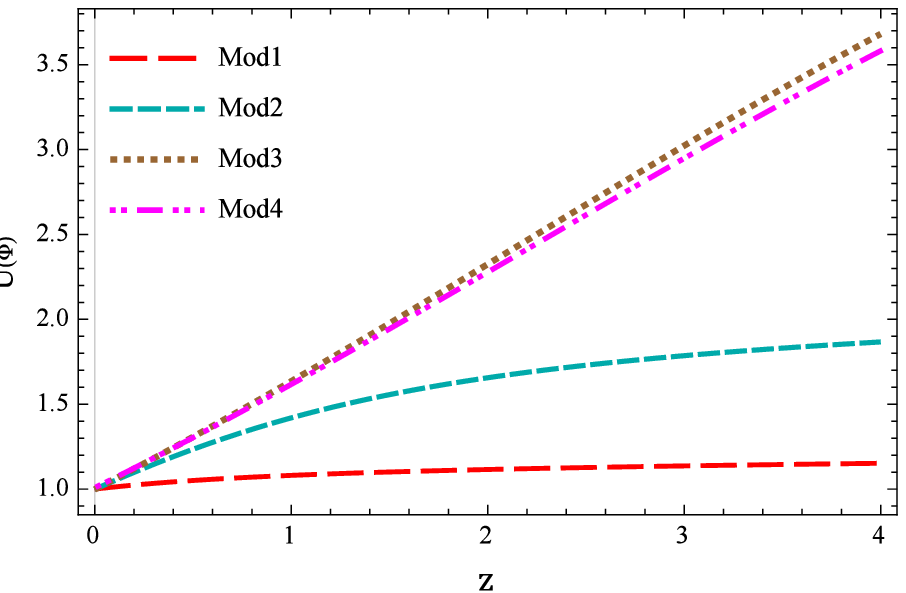}
 \includegraphics[width=8cm]{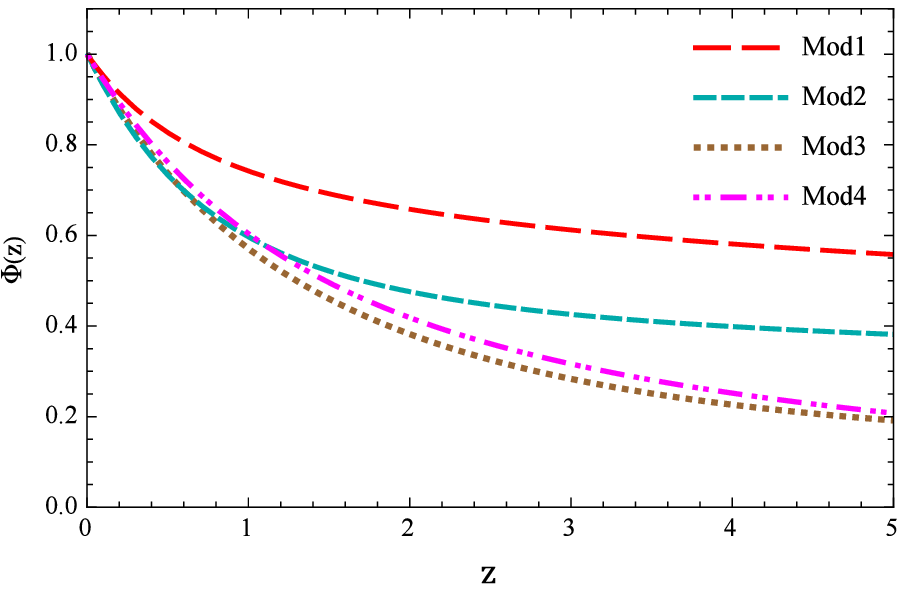}
 \caption{The redshift evolution of $1/F(\Phi)$ (top panel), the Ratra-Peebles potential $U(\Phi)$ (middle panel) and
 the scalar field (bottom panel) for the different cosmological models in ST cosmologies considered in
 Table~(\ref{tab:model}). The pink dot-dotted-dashed curve stands for the negative coupling constant, ($\xi<0$,
 model 4). The red long-dashed curve indicates the positive coupling $\xi=0.123$ (model 1), the green short-dashed
 curve corresponds to the positive coupling $\xi=0.088$ (model 2) and the brown dotted line represents the
 minimally coupled model $\xi=0$ (model 3).}
 \label{fig:F_phi}
\end{figure}

In figure~(\ref{fig:F_phi}), we show the evolution of $1/F(\Phi)$ (top panel), $U(\Phi)$ (middle panel) and $\Phi$
(bottom panel), as a function of redshift $z$ for the different cosmological models presented in
Table~(\ref{tab:model}). Line styles and colours for each model have been indicated in the caption. While $1/F(\Phi)$
is almost constant at high redshifts and its variation with cosmic $z$ is very small, it changes rapidly at low
redshifts and reaches one at present time. The constant line $1/F(\Phi)=1$ represents model (3), the minimally coupled
quintessence model with $\xi=0$, as expected. In the next section we will see that the effective gravitational constant
in ST cosmologies is proportional to $1/F(\Phi)$. In middle panel of figure~(\ref{fig:F_phi}) we show the evolution of
the Ratra-Peebles potential for the different models of Table~(\ref{tab:model}). In all the models, the potential
decays from larger values at higher redshift to one at the present time. In fact we fix the potential to be one at the
present time by setting $\Phi_0=1$ and $M=1$. We see that for the non-minimally coupled cases, independently of the
sign of $\xi$, the potential is weaker compared to the minimally coupled case in GR. A larger value of the coupling
parameter $\xi$ causes stronger deviations from the potential in GR gravity. Finally, in the bottom panel, we present
the redshift evolution of the scalar field $\Phi$ for different models. We see that for all models the scalar field
increases with redshift from its small initial value at $a_i=10^{-4}$ to $\Phi_0=1$ at the present time.

\begin{figure}
 \centering
 \includegraphics[width=8cm]{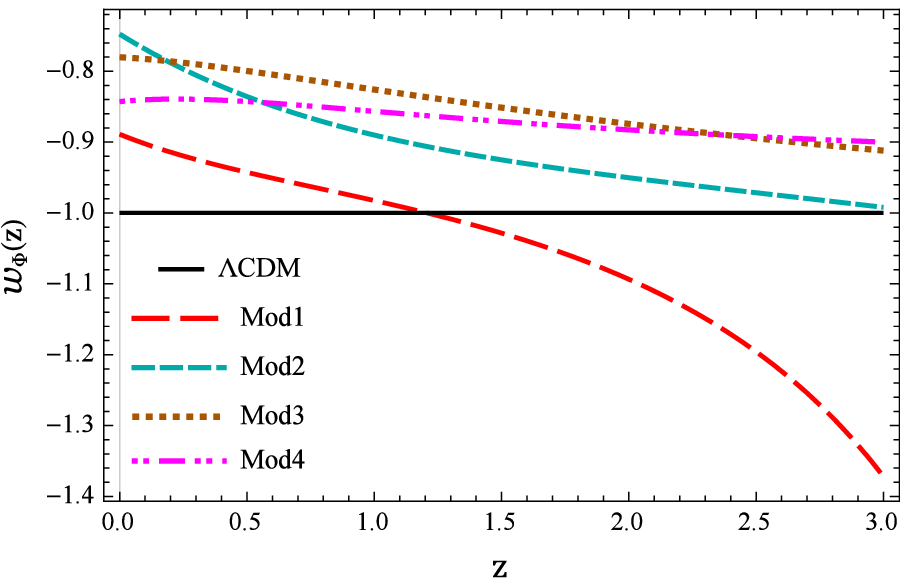}
 \includegraphics[width=8cm]{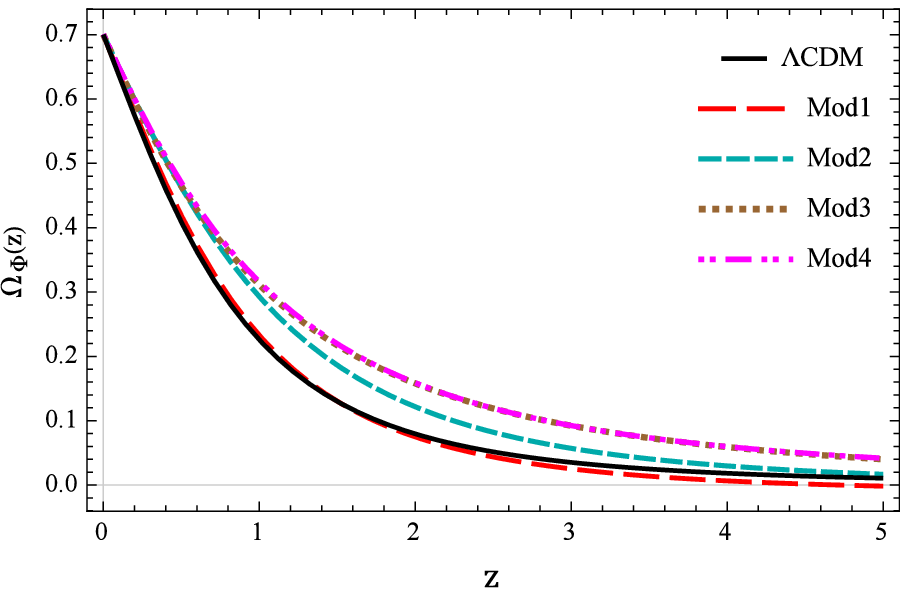}
 \includegraphics[width=8cm]{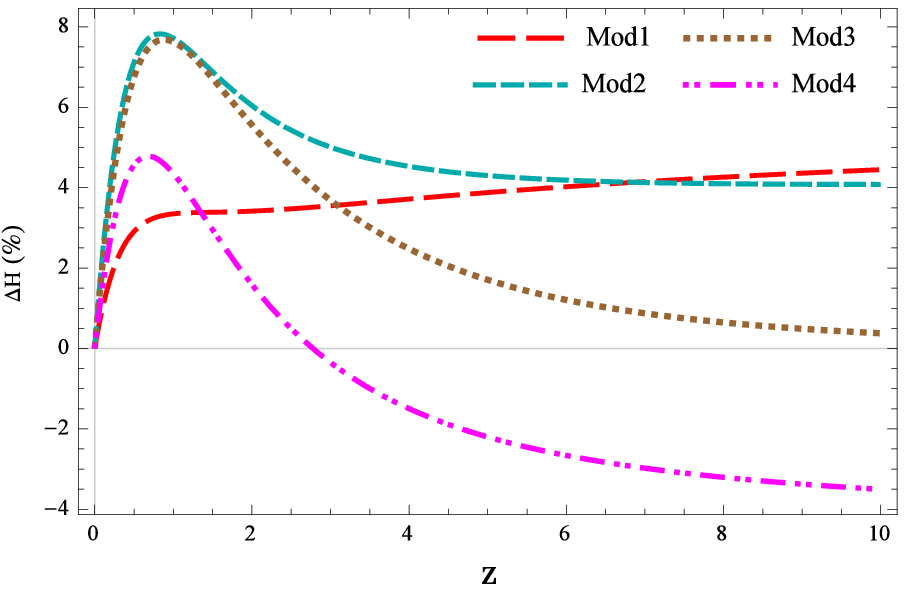}
 \caption{Top panel: The redshift evolution of the equation of state $w_{\Phi}$. Middle panel: the energy density of
 the minimally and non-minimally coupled quintessence models ($\Omega_{\rm \Phi}$) and $\Omega_{\Lambda}$ for the
 reference $\Lambda$CDM model (black solid curve). Bottom panel: the relative difference of the Hubble parameter in
 the framework of different minimally and non-minimally coupled quintessence models ($\Delta H(z)$) relative to the
 one of the reference $\Lambda$CDM model $H_{\rm \Lambda CDM}$. Line styles and colours are the same as in
 figure~(\protect{\ref{fig:F_phi}}).}
 \label{fig:background}
\end{figure}

We now calculate the redshift evolution of the background cosmological parameters: the equation of state of the scalar
field, $w_{\rm \Phi}=p_{\rm \Phi}/\rho_{\rm \Phi}$, the energy density parameter of the scalar field,
$\Omega_{\rm \Phi}$ and the Hubble expansion rate $H$, for the different cosmological models considered in this work.
As well known, these quantities can describe the background evolution of the universe. Moreover, the linear growth
rate of structures strongly depends on the background evolution. In figure~(\ref{fig:background}), using
equations~(\ref{eq:energy_density}) and~(\ref{eq:pressure}), we first show the evolution of $w_{\rm \Phi}$ as a
function of the redshift $z$ (top panel). We then show the redshift evolution of the energy density $\Omega_{\rm \Phi}$
for our selected models and $\Omega_{\Lambda}$ for concordance $\Lambda$CDM model (middle panel). We finally present
the evolution of the fractional difference of the Hubble parameter $\Delta H(z)$ for these models relatively to the
concordance $\Lambda$CDM model $H_{\rm\Lambda CDM}(z)$ in the bottom panel of figure~(\ref{fig:background}). We
refer to the caption for the different colours and line styles. In the case of a positive coupling ($\xi>0$), the
phantom regime ($w<-1$) can be achieved at high redshifts, while models with $\xi<0$ and $\xi=0$ remain in the
quintessence regime ($w>-1$). The amount of DE for different minimally and non-minimally coupled quintessence models as
well as the $\Lambda$CDM model have been presented in the middle panel. For all the models the amount of DE is
negligible at high redshifts meaning that all the models reduce to an EdS Universe at early times. The evolution of
$\Delta H(z)$ for different models is shown in the bottom panel. As it appears clear, at low redshifts, all the models
give $\Delta H(z)>0$ indicating that all of them have larger Hubble parameter than the $\Lambda$CDM universe. We see
that at high redshifts the Hubble parameter becomes smaller than the reference one only for a negative coupling
constant (model 4). In the case of the minimally coupled quintessence (model 3), the Hubble parameter tends to the
fiducial value in the $\Lambda$CDM model at high redshifts and differences become negligible. This result is
interesting, because in this model there is no direct coupling between the scalar field and gravity and therefore DE
models mimics the cosmological constant at high redshifts. In the case of non-minimally coupled quintessence models
with positive coupling $\xi>0$ (models 1 and 2), differences with respect to the $\Lambda$CDM model are at most $4\%$
at high redshifts for both models, independently of the exact value of $\xi$. We conclude that in ST theories of
gravity, the coupling between the scalar field and the gravitational sector causes differences in the Hubble expansion
rate with respect to the $\Lambda$CDM universe also at high redshifts. More in detail, at high redshifts differences
are positive (negative) for positive (negative) values of the coupling constant $\xi$. In the next section, we will
see that differences of the Hubble parameter together with the time evolution of the gravitational constant lead
to differences in the growth of matter perturbations in ST theories with respect to the standard general relativistic
models.

\section{Perturbations in ST gravities}\label{sect:perturbations}
Let us start with the perturbed FRW metric in the Newtonian gauge
\citep{Boisseau2000,EspositoFarese2001,BuenoSanchez2010}

\begin{equation}
 ds^2=-(1+2\phi)dt^2+a^2(1-2\psi)\delta_{ij}~dx^i~dx^j\;,\label{pertmet}
\end{equation}
where $\phi$ and $\psi$ are the linear gravitational potentials. In the framework of GR and in the absence of
anisotropic stresses, it is obvious to have $\phi=\psi$. In ST theories of gravity, this is no longer true and the
two potentials can be related to each other as \citep{Boisseau2000,EspositoFarese2001,BuenoSanchez2010}:

\begin{equation}
 \phi=\psi-\frac{F_{,\Phi}}{F}\delta\Phi\;, \label{psiphicon}
\end{equation}
where $F_{,\Phi}=dF/d\Phi$. For $F=1$ (as in GR gravity), equation~(\ref{psiphicon}) gives $\phi=\psi$ as expected.
The general relativistic equations for the perturbations of $\phi$, $\psi$ and the scalar field $\Phi$ in ST theories
of gravity have been studied in details by \cite{EspositoFarese2001,Hwang2005,Copeland2006,BuenoSanchez2010}.

The linear evolution of non-relativistic matter density perturbations and scalar field perturbations in the framework
of ST theory of gravity can be written as follows \citep{Copeland2006,BuenoSanchez2010}:

\begin{equation}
 \ddot{{\delta}}_{\rm m}+2H\dot{{\delta}}_{\rm m}+\frac{k^2}{a^2}
 \left(\psi-\frac{F_{,\Phi}}{F}\delta\Phi\right)-3(\ddot{\psi}+2H\dot{\psi})=0 \;,
 \label{matperteq}
\end{equation}
where $k$ is the wave number of the perturbations mode. Since in this work we deal with the SCM formalism with
perturbations well inside the horizon scale, in the following we will limit ourselves to the study of perturbations of
matter and scalar fields on scales much smaller than the Hubble scale ($k/a>>H$). On these scales, the scalar field
perturbations can be obtained as \citep{BuenoSanchez2010}:

\begin{equation}
 \delta\Phi \simeq (\phi-2\psi) F_{,\Phi}\;. \label{shdphi1}
\end{equation}
Using equation~(\ref{psiphicon}), we have
\begin{equation}
 \delta \Phi\simeq -\psi \frac{F F_{,\Phi}}{F+F_{,\Phi}^2}\;, \label{shdphi2}
\end{equation}
showing that the perturbations of non-minimally coupled scalar fields in ST gravity are scale independent on
sub-horizon scales.
For minimally coupled scalar field models in GR, since $F=1$ and consequently $F_{,\Phi}=0$, it is trivial to see that
$\delta\Phi=0$. Hence scalar field perturbations on sub-horizon scales are negligible for minimally coupled
quintessence models in GR gravity
\citep[see also][]{Unnikrishnan2008a,Jassal2009,BuenoSanchez2010,Jassal2010,Jassal2012}.

It has been shown that on sub-horizon scales, the energy density perturbation of the non-minimally coupled DE,
$\delta\rho_\Phi$, can be expressed as a function of scalar field perturbation $\delta\Phi$ as follows
\citep{BuenoSanchez2010}:
\begin{equation}
 \delta \rho_\Phi\simeq -\frac{k^2}{a^2} F_{,\Phi}\delta \Phi =
 \frac{k^2}{a^2}\psi \frac{F\;F_{,\Phi}^2}{F+F_{,\Phi}^2}\;. \label{shphipert}
\end{equation}
Also, matter density perturbations are \citep{BuenoSanchez2010}:
\begin{equation}
 \delta\rho_{\rm m}\simeq-\frac{k^2}{a^2}\psi F\left(\frac{F_{,\Phi}^2}{F+F_{,\Phi}^2}+2 \right)\;.
 \label{shmatterpert}
\end{equation}
Hence, the ratio $\frac{\delta\rho_\Phi}{\delta \rho_m}$ on sub-horizon scales is given by
\begin{equation}
 \frac{\delta\rho_\Phi}{\delta\rho_{\rm m}}=\frac{\delta_\Phi}{\delta_{\rm m}}\simeq
 -\frac{F_{,\Phi}^2}{3F_{,\Phi}^2+2F}\;. \label{shdrat}
\end{equation}
Using equations~(\ref{shphipert}) and~(\ref{shdrat}), we can eliminate the term $\frac{k^2}{a^2}\psi$ from
equation~(\ref{matperteq}) and being the study of the perturbations limited to the sub-horizon scales, we can also
ignore the time derivatives of the potential $\psi$ in equation~(\ref{matperteq})
\citep[see also][]{Boisseau2000,BuenoSanchez2010}. In this case, it is easy to obtain the linear evolution of matter
overdensities on sub-horizon scales within ST gravities as \citep[see also][]{Boisseau2000}
\begin{eqnarray}
 \ddot{\delta}_{\rm m}+2H\dot{\delta}_{\rm m}-4\pi G_{\rm eff}\rho_{\rm m}\delta_{\rm m}=0 \;,\label{matperteq2}
\end{eqnarray}
where the effective gravitational constant $G_{\rm eff}$ reads:
\begin{equation}
 G_{\rm eff}^{(p)} = \frac{G_{\rm N}}{F}\left(\frac{2F+4F_{,\Phi}^2}{2F+3F_{,\Phi}^2}\right)\;, \label{Geff}
\end{equation}
where ($p$) represents the clustering non-minimally coupled quintessence models. We see that
equation~(\ref{matperteq2}) is scale independent (as in the GR case) but the Newtonian gravitational constant
$G_{\rm N}$ is replaced by the effective gravitational constant $G_{\rm eff}$ as given by equation~(\ref{Geff}). In
fact, the effect of scalar field perturbations $\delta\Phi$ is included in the definition of $G_{\rm eff}$.

\begin{figure}
 \centering
 \includegraphics[width=8cm]{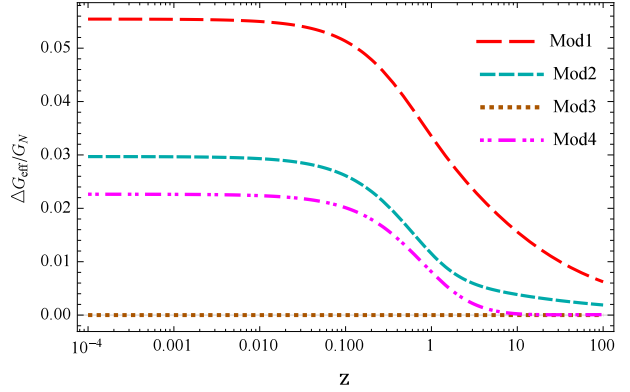}
 \caption{The redshift variation of the difference between the effective gravitational constant defined in clustering
 non-minimally coupled quintessence models $G_{\rm eff}^{(p)}$ and the same quantity in homogeneous non-minimally
 coupled quintessence models $G_{\rm eff}^{(h)}$, divided by Newtonian gravitational constant $G_{\rm N}$ as
 $\Delta G_{\rm eff}/G_{\rm N}=(G_{\rm eff}^{(p)}-G_{\rm eff}^{(h)})/G_{\rm N}$. Line styles and colours are the same
 as in figure~(\protect{\ref{fig:F_phi}}).}
 \label{fig:G_effective}
\end{figure}
In the case of homogeneous non-minimally coupled models where we ignore the perturbations of scalar field
($\delta{\Phi}=0$), it is easy to show that:
\begin{equation}\label{eq:k_over_a}
 \frac{k^2}{a^2}\psi=-4\pi\rho_{\rm m}\delta_{\rm m}\frac{1}{F}\left(\frac{2F+2F_{,\Phi}^2}{2F+3F_{,\Phi}^2}\right)\;.
\end{equation}
Inserting equation~(\ref{eq:k_over_a}) into equation~(\ref{matperteq}) and ignoring the time derivatives of $\psi$ in
the sub-Hubble scale regimes, we again obtain equation~(\ref{matperteq2}) for the evolution of matter overdensity in
homogeneous non-minimally quintessence models with $\delta{\rm \Phi}=0$, but in this case the effective gravitational
constant $G_{\rm eff}$ is reduced to \citep[see also][]{Pace2014}:
\begin{equation}
 G_{\rm eff}^{(h)}=\frac{G_{\rm N}}{F}\left(\frac{2F+2F_{,\Phi}^2}{2F+3F_{,\Phi}^2}\right)\;, \label{Geff2}
\end{equation}
where ($h$) indicates the homogeneous scenarios. In figure~(\ref{fig:G_effective}) we show the redshift variation of
the difference between $G_{\rm eff}^{(p)}$ and $G_{\rm eff}^{(h)}$ divided by the Newtonian gravitational constant
$G_{\rm N}$, $\Delta G_{\rm eff}/G_{\rm N}=(G_{\rm eff}^{(p)}-G_{\rm eff}^{(h)})/G_{\rm N}$, where
$G_{\rm eff}^{(p)}$ is the effective gravitational constant for clustering non-minimally coupled models and
$G_{\rm eff}^{(h)}$ is for the homogeneous case, respectively, defined in equations~(\ref{Geff} and~\ref{Geff2}).
In the case of model (3), we see that the relative difference $\Delta G_{\rm eff}/G_{\rm N}=0$. This result is
expected, since there are no perturbations of scalar field in this model and the effective gravitational constant
is given by equation~(\ref{Geff2}). In the case of models (1, 2 and 4), the non-minimally coupled models,
we see that $\Delta G_{\rm eff}/G_{\rm N}>0$ at low redshifts and approaches zero at high redshifts. Since the
difference between $G_{\rm eff}^{(p)}$ and $G_{\rm eff}^{(h)}$ at high redshifts is zero, we conclude that the
perturbations of the scalar field, which directly influence the effective gravitational constant, can be neglected at
early times. On the other hand, at low redshifts, since $\Delta G_{\rm eff}>0$, we can say that the effective
gravitational constant defined in clustering non-minimally quintessence models, $G_{\rm eff}^{(p)}$, is bigger than the
corresponding quantity defined in homogeneous non-minimally quintessence models. Moreover in model (1), where the
strength of non-minimally coupling between scalar field and gravity is the largest, the difference between
$G_{\rm eff}^{(p)}$ and $G_{\rm eff}^{(h)}$ is the largest (roughly $5.5\%$) among the other models.
On the basis of the above discussion, it is easy to see that due to the different behaviours of the definition of the
effective gravitational constant $G_{\rm eff}$ in clustering and homogeneous versions of non-minimally coupled models,
matter overdensity will evolve differently whether the scalar field perturbations are taken into account or not.
In the next section we solve equation~(\ref{matperteq2}) in order to follow the evolution of matter overdensities on
sub-Hubble scales in the linear regime for these two different classes of models. We also compare these models with
the minimally coupled quintessence model in GR gravity.

\section{Growth of overdensities in ST gravities}\label{sect:growth_structures}
In this section we study the growth of structures on sub-Horizon scales in the framework of ST gravity for different
homogeneous and clustering models. In particular, we evaluate the linear evolution of perturbations and the growth
factor in section~(\ref{sect:growth}), the non-linear evolution and the spherical collapse parameters in
section~(\ref{sect:scm}) and the mass function for virialised haloes in section~(\ref{mass_function}).

\subsection{Growth factor}\label{sect:growth}
Here we follow the linear growth of perturbations of non-relativistic dust matter and density perturbations of the
scalar field by solving equations~(\ref{shdrat}) and~(\ref{matperteq2}). We remind the reader that in the case of
clustering quintessence models where the effects of scalar field perturbations are taking into account, the effective
gravitational constant $G_{\rm eff}$ is given by equation~(\ref{Geff}), while in the case of homogeneous models, we
adopt equation~(\ref{Geff2}). It is also obvious, in the case of minimally coupled quintessence models ($\xi=0$), that
we have $G_{\rm eff}=G_{\rm N}$ as expected.

\begin{figure}
 \centering
 \includegraphics[width=8cm]{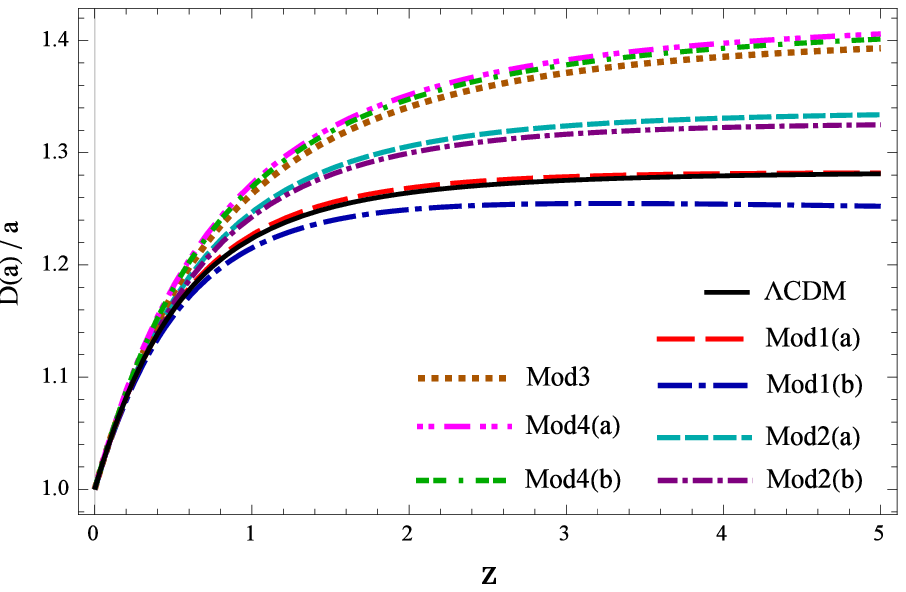}
 \includegraphics[width=8.5cm]{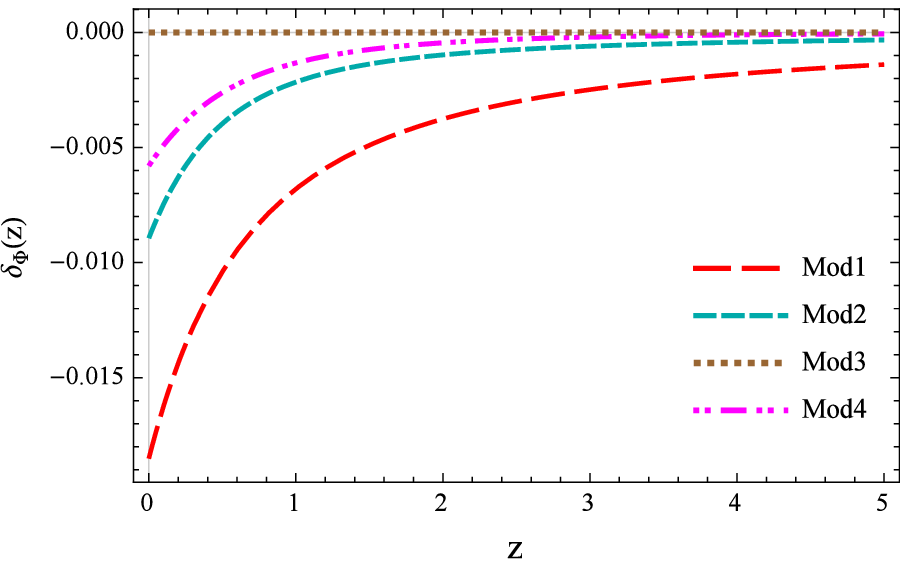}
 \caption{Top panel: Redshift evolution of the growth factor normalized to one at the present time divided by the
 scale factor $a$ in terms of the cosmic redshift $z$ for different cosmological models considered in
 Table~(\ref{tab:model}). For all the models, label (a) represents homogeneous models and label (b) indicates the
 clustering cases. The red long dashed (blue dotted-long-dashed) curve represents homogeneous (clustering) models with
 coupling parameter $\xi=0.123$ (model 1). The cyan dashed (violet dotted-dashed) curve represents the homogeneous
 (clustering) model with $\xi=0.088$ (model 2). The brown dotted curve indicates the minimally coupled quintessence
 model with $\xi=0.00$ (model 3). The pink dashed-dot-dotted (green dot-dashed-dashed) curve represents the
 homogeneous (clustering) model with $\xi=-0.087$ (model 4). The reference $\Lambda$CDM model is shown by a black solid
 curve for comparison. Bottom panel: Evolution of the density perturbations of the scalar field $\delta_{\Phi}$
 according to equation~(\ref{shdrat}). Line styles and colours are as in figure~(\ref{fig:F_phi}).}
 \label{fig:growth_factor}
\end{figure}

In figure~(\ref{fig:growth_factor}) we show the linear growth factor $D_{+}(a)=\delta_{\rm m}(a)/\delta_{\rm m}(a=1)$
divided by the scale factor ($D_{+}(a)/a$) as a function of redshift $z$ for the different cosmological models studied
in this work (top panel). In the bottom panel, we show the evolution of the density perturbations of dark energy
$\delta_{\rm \Phi}$ in terms of the redshift $z$. In all cases, label (a) represents homogeneous models and label (b)
indicates the clustering case. Line style and colours have been described in the caption of the figure.
While model (3), the minimally coupled case, has a larger growth factor compared to the reference $\Lambda$CDM universe
at high redshifts, we see that the non-minimally coupled models with negative (positive) coupling constant $\xi$ have a
higher (lower) growth factor with respect to the minimally coupled model. For all the models, the decrease of the
growth factor at lower redshifts is due to the fact that at late times DE dominates the energy budget of the universe
and suppresses the amplitude of perturbations. Also, the constant behaviour at high redshifts shows that at early times
the effects of DE are negligible for all the models. In the non-minimally coupled cases, we conclude that the growth
factor is smaller for clustering models compared to the homogeneous case. The difference between the growth factor of
clustering and homogeneous models is more pronounced for higher values of the coupling constant $\xi$. In fact the
decrease of the growth factor in clustering dark energy models compared to the homogeneous models is due to the fact
that the density perturbation of the dark energy field is always negative ($\delta_{\Phi}<0$) as shown in the bottom
panel of figure~(\ref{fig:growth_factor}). In addition, the negative value of $\delta_{\Phi}$ is due to the minus sign
in equation~(\ref{shdrat}) and one can see that clustering dark energy models can reproduce void DE structures
\citep[see also][]{Dutta2007,Unnikrishnan2008a,Jassal2009,Mainini2009,BuenoSanchez2010,Jassal2012,Villata2012}.

\subsection{SCM parameters}\label{sect:scm}
The linear overdensity parameter $\delta_{\rm c}$ and the virial overdensity $\Delta_{\rm vir}$ are the two main
quantities characterising the SCM. In this section we evaluate them in the framework of ST cosmologies.

The linear overdensity $\delta_{\rm c}$ is an important quantity in the Press Schechter formalism
\citep{Press1974,Bond1991,Sheth2002} and the virial overdensity $\Delta_{\rm vir}$ is used to determine the size of
virialised haloes.

To derive the time evolution of the linear overdensity parameter $\delta_{\rm c}$, we use the following non-linear
evolution equation \citep{Pace2014}
\begin{eqnarray}
 \ddot{\delta}_{\rm m}+2H\dot{\delta}_{\rm m}-\frac{4}{3}\frac{\dot{\delta}_{\rm m}^2}{1+\delta_{\rm m}} -
 4\pi G_{\rm eff}\rho_{\rm m}\delta_{\rm m}=0 \;,\label{matperteq3}
\end{eqnarray}
where in the case of homogeneous models $G_{\rm eff}$ is given by
equation~(\ref{Geff2}) and in the case of clustering models we use
equation~(\ref{Geff}). Remember that in the case of the minimally
coupled model (model (3) in our analysis), the effective
gravitational constant $G_{\rm eff}$ reduces to the constant
Newtonian gravitational constant $G_{\rm N}$ in the GR limit. It is
important to note that the evolution of matter perturbations in ST
cosmologies both in the linear (equation~\ref{matperteq2}) and
non-linear (equation~\ref{matperteq3}) regime are coupled to the
perturbations of the scalar field $\delta_{\rm \Phi}$ through the
effective gravitational constant $G_{\rm eff}$ introduced in
equations (\ref{Geff}) for clustering non-minimally coupled dark
energy models. We should emphasize that although the functional form
of equations (\ref{matperteq2} and \ref{matperteq3}) for the
evolution of matter perturbations is identical to that obtained in
standard GR once $G_{\rm N}$ is replaced by $G_{\rm eff}$,
differences between the two gravitational models are deeper. In fact
$G_{\rm eff}$ in ST theory is dynamical and evolves in time, while
$G_{\rm N}$ is constant in GR. Moreover, the perturbations of the
scalar field $\delta_{\rm \Phi}$ affect directly the evolution of
$G_{\rm eff}$ via equation (\ref{Geff}). This feature of ST theory
completely determine the behaviour of the growth of matter
perturbations in ST cosmologies with respect to the standard
perturbations growth in GR models. Said this, we follow the general
approach outlined in \cite{Pace2010,Pace2012,Pace2014} to determine
the linear overdensity $\delta_{\rm c}$ and virial overdensity
$\Delta_{\rm vir}$, to which we refer for an in depth description of
the procedure.

\begin{figure}
 \centering
 \includegraphics[width=8cm]{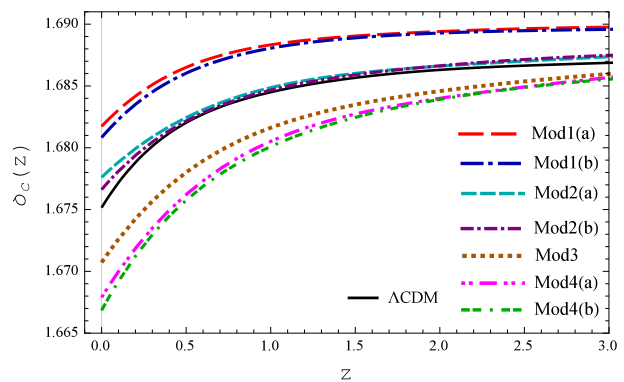}
 \caption{Linear overdensity parameter $\delta_{\rm c}$ as a function of the collapse redshift $z_{\rm c}$ for
 different minimally and non-minimally coupled models considered in this work. Line styles and colours
 are as in figure~(\ref{fig:growth_factor}).}
 \label{fig:delta_c}
\end{figure}

In figure~(\ref{fig:delta_c}), we show the evolution of $\delta_{\rm c}$ as a function of the collapse redshift
$z_{\rm c}$ for the models presented in Table~(\ref{tab:model}). We refer to the caption for line styles and colours.
In analogy to the previous section, label (a) represents homogeneous models and label (b) indicates clustering models.
We see that at low redshifts, where DE dominates the energy budget of the Universe, $\delta_{\rm c}$ for clustering
models is smaller than that obtained in the homogeneous case. In fact the perturbations of the scalar field due to the
non-minimally coupling between the scalar field and the Ricci scalar lowers the value of $\delta_{\rm c}$ at low
redshifts. This results is expected, since we showed that the density perturbations of scalar fields due to the
non-minimally coupling are always negative (see bottom panel of figure~\ref{fig:growth_factor}). Furthermore, one
can see that for positive (negative) coupling constant $\xi$ the value of $\delta_{\rm c}$ is larger (smaller)
compared to the minimally coupled model (e.g., model 3).

In addition to $\delta_{\rm c}$, the other important parameter in the SCM is the virial overdensity $\Delta_{\rm vir}$.
The size of spherically symmetric haloes can be well defined by the virial overdensity parameter. The virial
overdensity is defined as $\Delta_{\rm vir}=\zeta (x/y)^3$, where $\zeta$ is the overdensity at the turn-around
redshift, $x$ is the scale factor normalised to the turn-around scale factor and $y$ is the ratio between the
virialised radius and the turn-around radius \citep{Wang1998}. In EdS cosmology, it is well known that $y=1/2$,
$\zeta\approx 5.6$ and $\Delta_{\rm vir}\approx178$ at any cosmic redshift. In DE cosmologies, $\Delta_{\rm vir}$
depends on the evolution of the DE sector and evolves with redshift. Moreover, according to whether DE takes part or
not into the virialization process, the quantity $y$ may be larger or smaller than $1/2$. Hence the virial overdensity
$\Delta_{\rm vir}$ can be affected by the clustering of DE \citep{Maor2005,Pace2014b,Malekjani2015}.

In standard cosmology, the virialization of pressureless dust matter and size of forming haloes are strongly affected
by the DE component \citep{Lahav1991,Wang1998,Mota2004,Horellou2005,Wang2005,Naderi2015} and also by DE perturbations
\citep{Abramo2007,Abramo2008,Abramo2009a,Batista2013,Pace2014b,Malekjani2015}.
In ST cosmologies we expect that the density perturbations of the scalar field derived from the non-minimally coupling
between the scalar field and Ricci scalar affect the evolution of the virial overdensity $\Delta_{\rm vir}$.

\begin{figure}
 \centering
 \includegraphics[width=7cm]{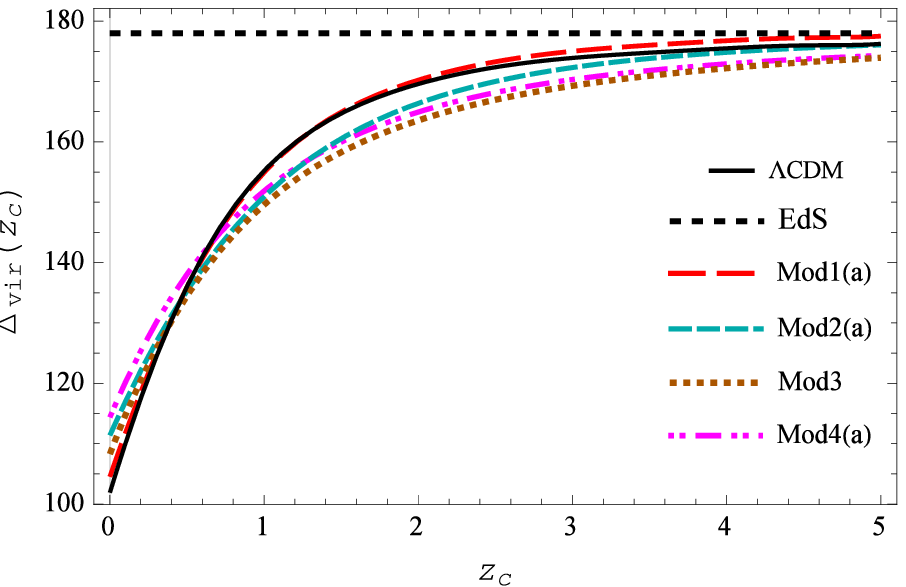}
 \includegraphics[width=7cm]{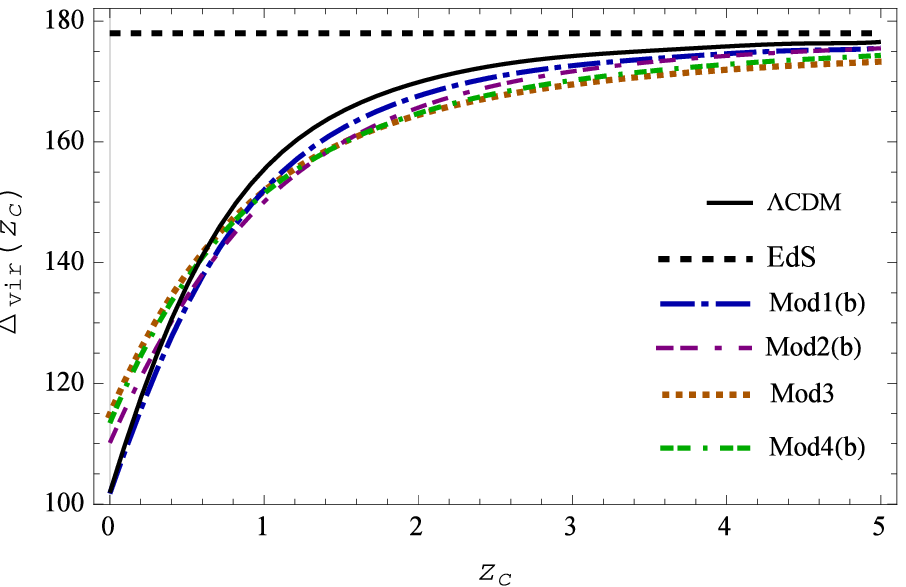}
 \caption{Virial overdensity parameter $\Delta_{\rm vir}$ as a function of collapse redshift $z$ for different
 models given in Table~(\ref{tab:model}). The top panel shows non-clustering models while the bottom panel represents
 the clustering case. Line styles and colours are as in figure~(\ref{fig:growth_factor}).}
 \label{fig:deltavir1}
\end{figure}

In figure~(\ref{fig:deltavir1}), we show the redshift evolution of the virial overdensity $\Delta_{\rm vir}$ for
different homogeneous (top panel) and clustering (bottom panel) models in ST theory. All clustering and homogeneous
quintessence models reach the fiducial value $\Delta_{\rm vir}\approx 178$ at high redshifts and decrease towards
smaller redshift values. This result is expected, since at high redshifts the effect of DE on the scenario of
structure formation is negligible and the EdS cosmology is recovered. A decrease in $\Delta_{\rm vir}$ with redshift
$z$ in ST cosmologies indicates that the quintessence sector in ST gravity prevents collapse and condensation of
overdense regions as it happens for DE models in standard GR gravity. For $a\approx 1$ the difference between
clustering and homogeneous quintessence models is limited. We see that in homogeneous models the virial overdensity
is slightly bigger than in the clustering case. Quantitatively speaking, the present value of $\Delta_{\rm vir}$ in
model (1a) is roughly $3\%$ larger than model (1b). In the case of homogeneous models (2a) and (4a), $\Delta_{\rm vir}$
is almost $1\%$ higher than clustering cases (2b) and (4b).

\begin{figure*}
 \begin{center}
  \includegraphics[width=7cm]{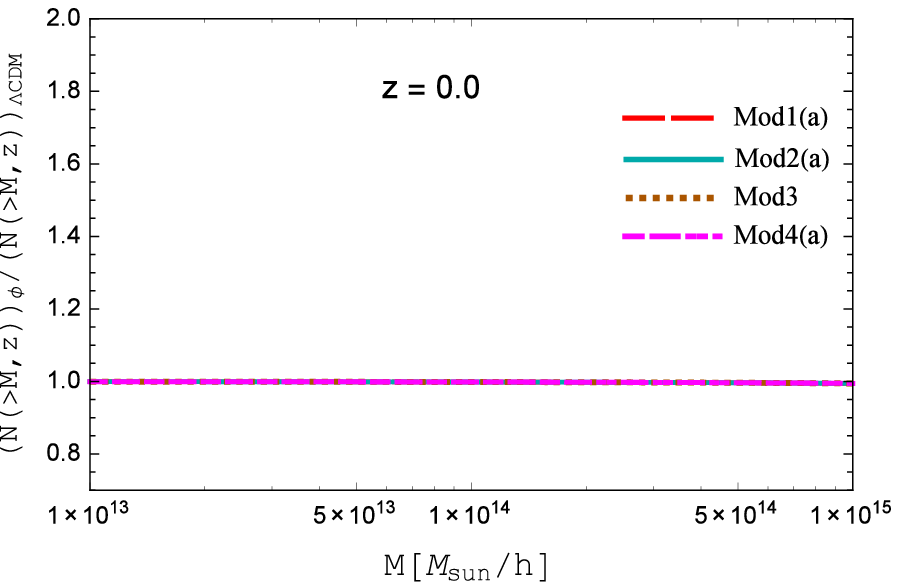}
  \includegraphics[width=7cm]{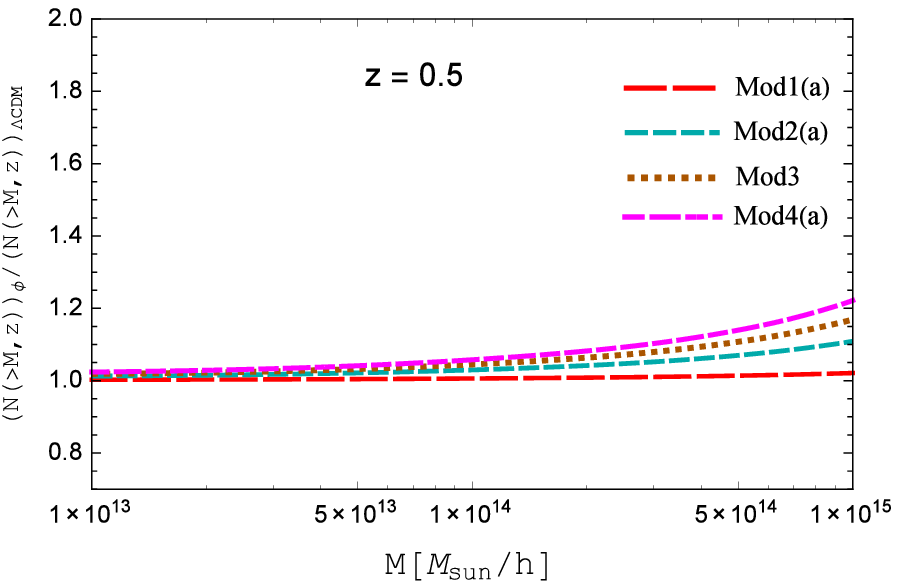}
  \includegraphics[width=7cm]{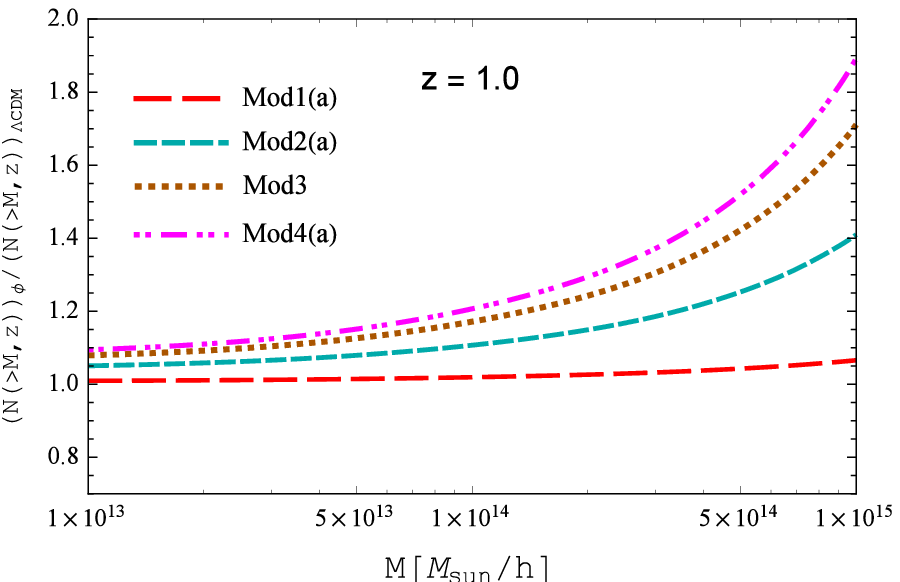}
  \includegraphics[width=7cm]{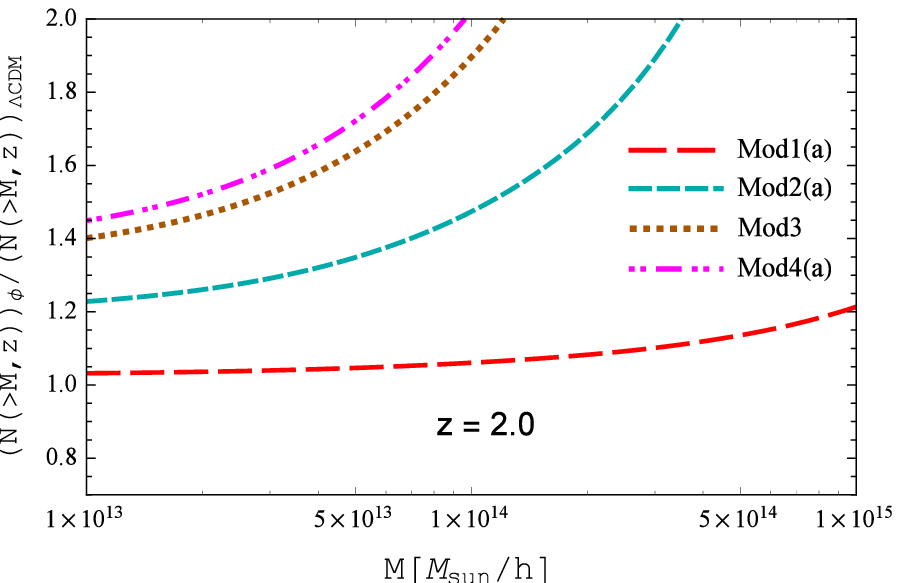}
  \caption{Ratio of the number of haloes above a given mass $M$ at $z=0$ (top left), $z=0.5$ (top right), $z=1.0$
  (bottom left) and $z=2.0$ (bottom right) between the minimally and the non-minimally coupled quintessence models
  and the concordance $\Lambda$CDM cosmology. Scalar field models have been assumed homogeneous. Line style and
  colours are as in figure~(\ref{fig:F_phi}).}
  \label{fig:mass_fun1}
 \end{center}
\end{figure*}

\subsection{Abundance of haloes in homogeneous quintessence models }\label{mass_function}
We now estimate the comoving number density of virialised haloes in a certain mass range. To this end, we adopt the
Press-Schechter formalism in which the abundance of haloes is described in terms of their mass \citep{Press1974}.
In the Press-Schechter formalism, the fraction of the volume of the Universe which collapses into an halo of mass $M$
at a given redshift $z$ is expressed by a Gaussian distribution function \citep{Press1974,Bond1991}. The comoving
number density of haloes with masses in the range of $M$ and $M+dM$ as a function of the redshift $z$ is given by
\begin{eqnarray}
 \frac{dn(M,w,z)}{dM}=\frac{\bar{\rho}_0}{M}\frac{d\nu(M,w,z)}{dM}f(\nu)\;,\label{eqn:PS1}
\end{eqnarray}
where $\bar{\rho}_0$ is the background density at the present time and
\begin{equation}
 \nu(M,w,z)=\frac{\delta_{\rm c}}{\sigma}\;,
\end{equation}
where $\sigma$ is the r.m.s. of the mass fluctuations in spheres containing the mass $M$. Generally, the parameters
$\delta_{\rm c}$ and $\sigma$ depend on the cosmological model and as a consequence also on the equation of state of
the dark energy component. Although, the standard mass function $f(\nu)$ presented in \cite{Press1974} can provide a
good estimate of the predicted number density of haloes, it fails by predicting too many low-mass haloes and too few
high mass objects \citep{Sheth1999,Sheth2002}. Hence we use a modified mass function proposed by
\cite{Sheth1999,Sheth2002}:

\begin{equation}
 f(\nu)= 0.2709\sqrt{\frac{2}{\pi}}\left(1+1.1096 \nu^{0.3}\right)
 \exp{\left(-\frac{0.707\nu^2}{2}\right)}\;.
 \label{eq:multiplicity_ST}
\end{equation}
Adopting a Gaussian density field, the amplitude of mass fluctuation $\sigma(M)$ is given by
\begin{equation}\label{eq:sigma}
 \sigma^2=\frac{1}{2\pi^2}\int_0^{\infty}{k^2P(k)}W^2(kR)dk\;,
\end{equation}
where $R$ is the radius of the overdense spherical region, $W(kR)$ is the Fourier transform of a spherical top-hat
filter and $P(k)$ is the linear power spectrum of density fluctuations \citep{Peebles1993}. The number density of
virialised haloes above a certain mass $M$ at collapse redshift $z$ is
\begin{equation}\label{eq:ndensity}
 n(>M,z)=\int_{M}^{\rm \infty}\frac{dn(z)}{dM^{\prime}}dM^{\prime}\;.
\end{equation}

We now compute the predicted number density of virialised haloes in the Press-Schechter formalism for homogeneous
models in ST cosmologies. In the case of homogeneous cosmologies we use equations~(\ref{eqn:PS1})
and~(\ref{eq:ndensity}) to determine the number density of virialised haloes. In this case the total mass of haloes is
defined by the pressureless matter perturbations. In order to calculate $\sigma^2$, we adopt the formulation presented
in \cite{Abramo2007,Naderi2015}. On the basis of latest observational results by the Planck Collaboration team
\citep{Planck2015_XIII}, we adopt the concordance $\Lambda$CDM model with the normalization of the matter power
spectrum $\sigma_8=0.815$.

In figure~(\ref{fig:mass_fun1}), we show the ratio of the predicted number of haloes above a given mass $M$ between the
homogeneous models in ST gravity and the concordance $\Lambda$CDM universe for different cosmic redshifts: $z=0$ (top
left panel), $z=0.5$ (top right panel), $z=1.0$ (bottom left panel) and $z=2.0$ (bottom right panel). Analogously to
previous figures, label (a) represents homogeneous models. We remind the reader that model (3) represents a minimally
coupled quintessence model. At $z\approx 0$, all the models produce approximately the same number of haloes over a
large mass range, however small differences take place at high masses.

At $z=0.5$, we see that all models including the $\Lambda$CDM one are still giving approximately the same number of
objects at low masses ($M\approx 10^{13}M_{\odot}/h$), while at high masses ($M\approx 10^{15}M_{\odot}/h$) the
differences between the models 2, 3, 4 and the $\Lambda$CDM one become more pronounced. However, differences are
negligible for model (1). Quantitatively speaking, at $z=0.5$, model (4) with negative coupling constant $\xi=-0.087$
has roughly $22\%$ more halos than the $\Lambda$CDM model. This value is roughly $11\%$ for model (2) and $17\%$ for
the minimally coupled quintessence case with $\xi=0$ (model 3). At higher redshifts, $z=1$ and $z=2$, we see that
differences in the halo numbers appear also at low masses. At high redshifts, all the models here investigated predict
a higher number of virialised haloes compared to the $\Lambda$CDM model. In particular, the number of objects in the
non-minimally coupled case with negative coupling parameter $\xi=-0.087$ (model 4) is larger than what is predicted for
a minimally coupled case (model 3). In the non-minimally coupled case with positive coupling constant $\xi=0.088$ and
$\xi=0.123$ (models 1 and 2) instead, the number of objects is smaller. Fractional differences in the number of
virialised haloes between different models and the $\Lambda$CDM model are presented in table~(\ref{tab:number1}).
Results are shown for three mass scales: $M>10^{13}M_{\odot}/h$, $M>10^{14}M_{\odot}/h$ and $M>10^{15}M_{\odot}/h$.

\begin{table}
 \centering
 \caption[]{Numerical results for the fractional difference of number of haloes between homogeneous minimally and
 non-minimally coupled quintessence models and the concordance $\Lambda$CDM model. These results are presented at
 four different redshifts: $z=0$, $z=0.5$, $z=1$ and $z=2$ for objects with $M>10^{13}M_{\odot}/h$ (low mass scale),
 $M>10^{14}M_{\odot}/h$ (intermediate mass scale) and $M>10^{15}M_{\odot}/h$ (high mass end).}
 \begin{tabular}{lcccc}
  \hline
  \hline
  Model 1(a) & $z=0$ & $z=0.5$ & $z=1$ & $z=2$\\
  \hline
  $M>10^{13}M_{\odot}/h$ & -0.1\% & 0.3\% & 0.9\% & 3\% \\
  $M>10^{14}M_{\odot}/h$ & -0.2\% & 0.6\% & 2\%   & 6\% \\
  $M>10^{15}M_{\odot}/h$ & -0.6\% & 2\%   & 7\%   & 21\% \\
  \hline
  \hline
  Model 2(a) & $z=0$ & $z=0.5$ & $z=1$ & $z=2$\\
  \hline
  $M>10^{13}M_{\odot}/h$ & -0.1\% & 1.2\% & 5\%  & 22\% \\
  $M>10^{14}M_{\odot}/h$ & -0.2\% & 3\%   & 11\% & 47\% \\
  $M>10^{15}M_{\odot}/h$ & -0.6\% & 11\%  & 40\% & - \\
  \hline
  \hline
  Model 3 & $z=0$ & $z=0.5$ & $z=1$ & $z=2$\\
  \hline
  $M>10^{13}M_{\odot}/h$ & -0.1\% & 2\%   & 8\%  & 39\% \\
  $M>10^{14}M_{\odot}/h$ & -0.2\% & 4\%   & 17\% & 89\% \\
  $M>10^{15}M_{\odot}/h$ & -0.5\% & 17\%  & 71\% & -\\
  \hline
  \hline
  Model 4(a) & $z=0$ & $z=0.5$ & $z=1$ & $z=2$\\
  \hline
  $M>10^{13}M_{\odot}/h$ & -0.1\% & 2.3\% & 9\%  & 43\% \\
  $M>10^{14}M_{\odot}/h$ & -0.2\% & 6\%   & 21\% & 101\% \\
  $M>10^{15}M_{\odot}/h$ & -0.5\% & 22\%  & 89\% & - \\
  \hline
 \label{tab:number1}
 \end{tabular}
 \begin{flushleft}
  \vspace{-0.5cm}
  {\small}
 \end{flushleft}
\end{table}

\begin{figure*}
 \begin{center}
  \includegraphics[width=7cm]{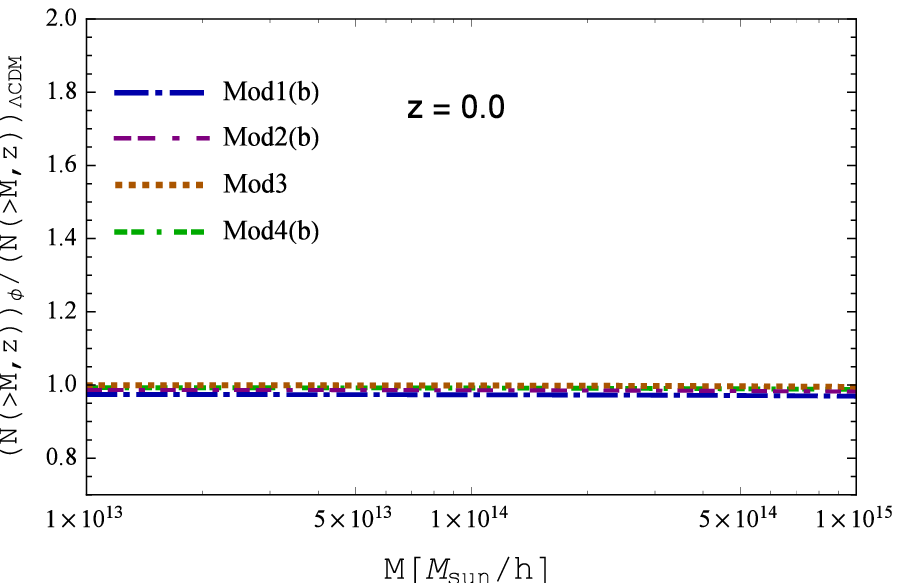}
  \includegraphics[width=7cm]{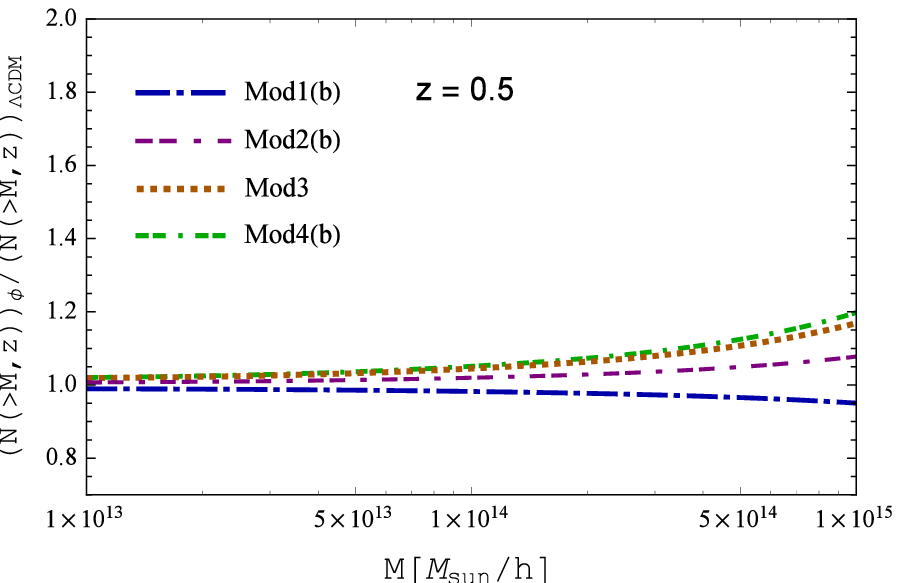}
  \includegraphics[width=7cm]{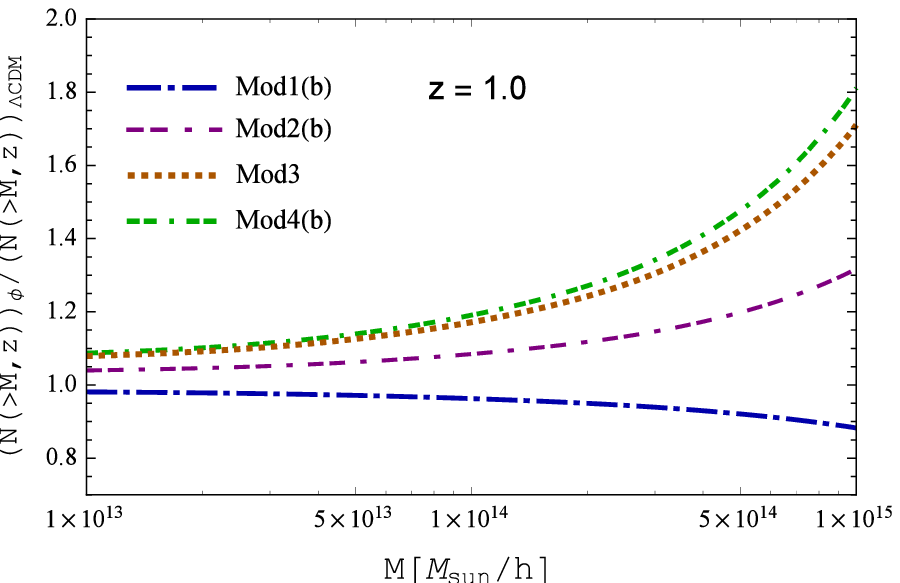}
  \includegraphics[width=7cm]{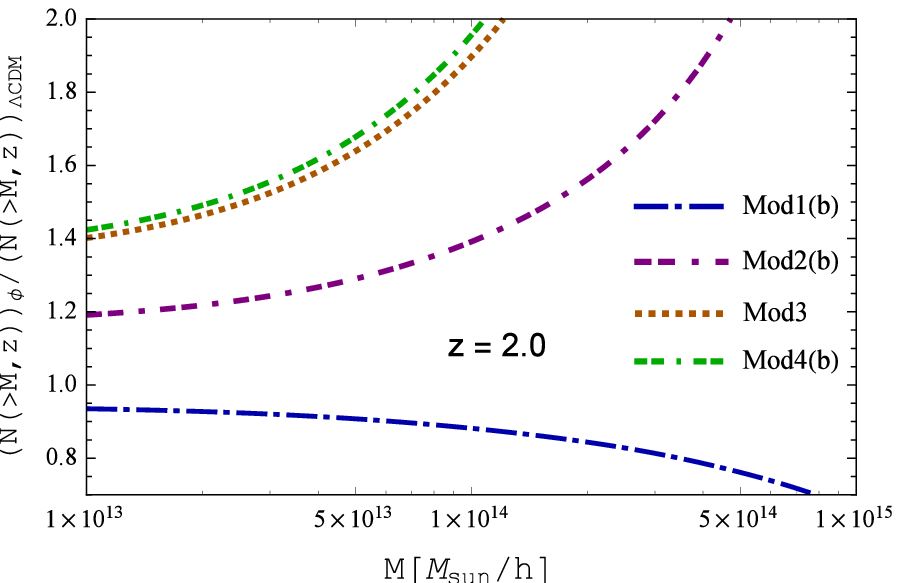}
  \caption{Ratio of the number of haloes above a given mass $M$ at $z=0$ (top left), $z=0.5$ (top right),
  $z=1.0$ (bottom left) and $z=2.0$ (bottom right) between the minimally and the non-minimally coupled scalar field
  models considered in this work and the concordance $\Lambda$CDM cosmology. Here we assume the clustering of scalar
  field models. Line style and colours are same as in figure~(\ref{fig:growth_factor}).}
  \label{fig:mass_fun2}
 \end{center}
\end{figure*}

\subsection{Abundance of haloes in clustering quintessence models}
As shown in section~(\ref{sect:scm}), the size and density of virialised haloes strongly depends on the background
dynamics and change in the presence of the scalar field perturbations. Hence in clustering models we should take into
account the contribution of the scalar field density perturbations to the total mass of the haloes
\citep[see also][]{Creminelli2010,Basse2011,Batista2013,Pace2014b,Malekjani2015}.
We follow the formulation presented before in DE cosmologies where the total mass of the haloes is affected by DE
perturbations \citep{Batista2013,Malekjani2015}. The quantity $\epsilon(z)=M_{\rm DE}/M_{\rm m}$, representing the
ratio of DE mass to matter mass, indicates how DE perturbations affect the total mass of virialised haloes. The mass
of DE is defined according to the contribution of DE perturbation. If we assume the top-hat density profile and fully
clustering DE, we have \citep[see also][]{Malekjani2015}

\begin{equation}
 \epsilon(z)=\frac{\Omega_{\rm \phi}(z)}{\Omega_{\rm m}(z)}\frac{\delta_{\rm de}(z)}{1+\delta_{\rm m}(z)}\;.
 \label{eq:epsilon1}
\end{equation}

\begin{figure}
 \centering
 \includegraphics[width=8cm]{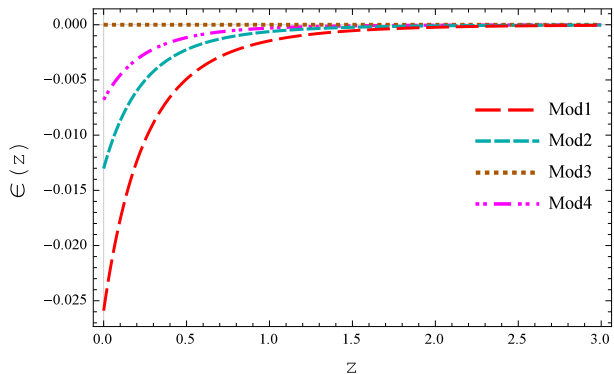}
 \caption{The redshift evolution of the ratio of DE mass to pressureless dark matter mass $\epsilon(z)$ calculated
 according to equation~(\ref{eq:epsilon1}) for different scalar field models considered in this work.
 For the non clustering minimally coupled scalar field case (model 3) we have $\epsilon=0$ as expected.
 Line style and colours are as in figure~(\ref{fig:F_phi}).}
\label{fig:epsilon}
\end{figure}

In figure~(\ref{fig:epsilon}), we show the redshift evolution of $\epsilon(z)$ on the basis of
equation~(\ref{eq:epsilon1}) for the different minimally and non-minimally coupled quintessence models considered in
this work. We see that for all the non-minimally coupled models with positive or negative coupling constant $\xi$, the
quantity $\epsilon$ is negative and for the minimally coupled case (model 3) $\epsilon$ is zero. This result is
expected, since $\delta_{\rm \Phi}$ for all non-minimally coupled models is negative and zero for model 3 (see also
figure~\ref{fig:growth_factor}). We also see that for $z\gtrsim1$, in all non-minimally coupled models $\epsilon$
approaches zero. This means that the contribution of the DE mass to the total mass of haloes is negligible at high
redshifts. This is nothing else than requiring an EdS behaviour in the past. Moreover, models with larger coupling
constant $\xi$ give a higher contribution of the DE mass to the total mass of haloes. We notice that in these models
the DE mass is negative and hence should be subtracted from the total mass of haloes.

Following \cite{Batista2013,Pace2014b,Malekjani2015}, in the presence of DE mass contribution, the definition of the
mass function in the Press-Schechter formalism is changed to
\begin{eqnarray}
 \frac{dn(M,w,z)}{dM}=\frac{\bar{\rho}_0}{M(1-\epsilon)}\frac{d\nu(M,w,z)}{dM}f(\nu)\;,
 \label{eqn:PS2}
\end{eqnarray}
where $f(\nu)$ is given by equation~(\ref{eq:multiplicity_ST}). We notice that in clustering models, density
perturbations of the scalar field change the function $f(\nu)$ via changing the quantities $\delta_{\rm c}$ and
$\sigma$. In figure~(\ref{fig:mass_fun2}), we show the predicted number of haloes calculated with the corrected mass
function in equation~(\ref{eqn:PS2}) for clustering models divided by those obtained for the concordance $\Lambda$CDM
model. In analogy to figure~(\ref{fig:mass_fun1}), we select four different redshifts $z=0$, $z=0.5$, $z=1$ and $z=2$.
As in the previous section, clustering models are labelled with the letter "(b)". Results for model (3) are similar to
those obtained before and presented in section~(\ref{mass_function}). We see that at $z\approx 0$, all models roughly
coincide with the $\Lambda$CDM case and differences are negligible. At $z=0.5$ differences with the $\Lambda$CDM model
are considerable at the high mass tail ($M>10^{15}M_{\odot}/h$). In particular, we see that for model (4b) the number
of haloes is roughly $20\%$ higher than that of the concordance $\Lambda$CDM model. This value is $17\%$ for model 3
and approximately $8\%$ for model (2b) and $-5\%$ in the case of model (1b). The numerical values for the fractional
difference of the number of haloes between the different clustering models considered in this work and the
$\Lambda$CDM model are presented in table~(\ref{tab:number2}). In analogy to table~(\ref{tab:number1}), the results
are reported for masses greater than $10^{13}M_{\odot}/h$, $10^{14}M_{\odot}/h$ and $10^{15}M_{\odot}/h$. At higher
redshifts, $z=1$ and $z=2$, we see that the predicted number of haloes calculated for clustering models deviates from
the $\Lambda$CDM expectation even at the low mass end of the halo mass function. Generally, by comparing the results
presented in tables~(\ref{tab:number1}) and~(\ref{tab:number2}), we conclude that the number of virialised haloes
estimated in clustering models is lower than that for homogeneous models. This result is due to the negative sign of
$\delta_{\Phi}$ and $\epsilon$ in clustering quintessence models. Differences between clustering and homogeneous models
are more pronounced at the high mass tail and at high redshifts, as expected. For example for model (1) where the
coupling constant is the largest ($\xi=0.123$), we see that at the high mass tail ($M>10^{15}M_{\odot}/h$ at $z=0$) the
homogeneous model (1a) produces only $2.5\%$ more objects than in the clustering case (model (1b)). This value is $7\%$
at $z=0.5$, roughly $19\%$ at $z=1$ and $55\%$ at $z=2$. For other quintessence models we have similar results.

\begin{table}
 \centering
 \caption[]{The fractional difference of number of haloes between clustering models in ST cosmologies and the
 concordance $\Lambda$CDM model. Results are shown at four different epochs: $z=0$, $z=0.5$, $z=1$ and $z=2$ for
 objects with $M>10^{13}M_{\odot}/h$, $M>10^{14}M_{\odot}/h$ and $M>10^{15}M_{\odot}/h$.}
 \begin{tabular}{lcccc}
  \hline
  \hline
  Model 1(b) & $z=0$ & $z=0.5$ & $z=1$ & $z=2$\\
  \hline
  $M>10^{13}M_{\odot}/h$ &-2.6\%&-3.2\% & -4.5\% & -6.4\% \\
  $M>10^{14}M_{\odot}/h$ &-2.7\%&-3.6\% &  -5.3\% & -12\% \\
  $M>10^{15}M_{\odot}/h$ &-3.1\%& -5\% & -11.8\% & -34\% \\
  \hline
  \hline
  Model 2(b) & $z=0$ & $z=0.5$ & $z=1$ & $z=2$\\
  \hline
  $M>10^{13}M_{\odot}/h$ &-1.4\%&0.6\% & 3.8\% & 19\% \\
  $M>10^{14}M_{\odot}/h$ &-1.5\%&1.9\% &8.5\% &39\% \\
  $M>10^{15}M_{\odot}/h$ &-1.9\%&7.7\% & 32\% & - \\
  \hline
  \hline
  Model 3 & $z=0$ & $z=0.5$ & $z=1$ & $z=2$\\
  \hline
  $M>10^{13}M_{\odot}/h$ &-0.1\%&2\% & 8\% & 39\% \\
  $M>10^{14}M_{\odot}/h$ &-0.2\%&4\% & 17\% & 89\% \\
  $M>10^{15}M_{\odot}/h$ &-0.5\%&17\% & 71\% & - \\
  \hline
  \hline
  Model 4(b) & $z=0$ & $z=0.5$ & $z=1$ & $z=2$\\
  \hline
  $M>10^{13}M_{\odot}/h$ &-0.8\%&2\% & 8.5\% &42\% \\
  $M>10^{14}M_{\odot}/h$ &-1\%&5\% & 19\% & 95\% \\
  $M>10^{15}M_{\odot}/h$ &-1.6\%&20\% & 81\% & - \\
  \hline
 \label{tab:number2}
 \end{tabular}
 \begin{flushleft}
  \vspace{-0.5cm}
  {\small}
 \end{flushleft}
\end{table}

\section{Conclusions}\label{sect:conclusions}
In the context of the spherical collapse model (SCM) we studied the non-linear growth of structures in scalar-tensor
(ST) cosmologies. In ST gravity, there is a non-minimally coupling between the scalar field and the Ricci scalar, the
so-called non-minimally coupled quintessence models. We first studied the background expansion history in ST theories.
We saw that in the case of positive non-minimally coupling parameter, $\xi>0$ (model 1), the equation of state of the
scalar field $w_{\rm \Phi}$ can achieve the phantom regime ($w_{\Phi}<-1$) at high redshifts, while in the case of
minimally coupled quintessence models (model 3), $w_{\rm \Phi}$ remains always in the quintessence regime
$-1<w_{\Phi}<-1/3$, as expected (see the top panel of figure~\ref{fig:background}). The redshift evolution of the
energy density $\Omega_{\rm \Phi}$ shows that all minimally and non-minimally coupled quintessence models considered
in this work reduce to an EdS universe at high redshifts where the dynamics of the universe can be well described by
the pressure-less matter component (see middle panel of figure~\ref{fig:background}). All quintessence models in ST
theories are characterised by $\Delta H(z)>0$ at low redshift, meaning that all minimally and non-minimally coupled
quintessence models have a larger Hubble parameter compared to the $\Lambda$CDM universe at low redshifts (see bottom
panel of figure~\ref{fig:background}).

We then followed the evolution of matter perturbations within the ST
gravity on sub-Hubble scales. In particular, we focused on the
scalar field perturbations due to the non-minimal coupling between
the scalar field and the Ricci scalar, the so called clustering
non-minimally coupled quintessence models. When we ignore the scalar
field perturbations and assume that the scalar field is important
only at the background level, the model is the so-called homogeneous
non-minimally coupled quintessence model. The equation describing
the evolution of matter overdensities in ST theories is similar to
the one found in general relativistic models, albeit in this case
the Newtonian gravitational constant $G_{\rm N}$ is replaced by the
effective gravitational constant $G_{\rm eff}$
(equation~\ref{Geff}). Since there is a difference between the
definition of $G_{\rm eff}$ in homogeneous (equation~\ref{Geff2})
and clustering (equation~\ref{Geff}) non-minimally coupled
quintessence models, we infer that the evolution of matter
perturbations on the basis of equation~(\ref{matperteq2}) differs
between clustering and homogeneous non-minimally coupled
quintessence models. We saw that for higher values of the coupling
parameter $\xi$, the effective gravitational constant $G_{\rm
eff}^{(p)}$ defined in clustering non-minimally quintessence model
is higher than the same quantity $G_{\rm eff}^{(h)}$ defined in
homogeneous non-minimally coupled quintessence models
(figure~\ref{fig:G_effective}). Hence we conclude that the scalar
field perturbations directly affect the effective gravitational
constant in ST cosmologies. In the case of minimally coupled
quintessence models ($\xi=0$) since the perturbations of scalar
field are vanishing, the difference between these two different
definitions of effective gravitational constant is zero. The linear
growth factor depends strongly on the sign of the coupling parameter
$\xi$, so that the growth factor for non-minimally coupled
quintessence with negative (positive) coupling $\xi$ is larger
(smaller) than in minimally coupled quintessence (top panel of
figure~\ref{fig:growth_factor}).

We showed that independently of the sign of the coupling parameter, the perturbations of the scalar field are always
negative (bottom panel of figure~\ref{fig:growth_factor}), so that the clustering non-minimally quintessence models
can reproduce the void DE structures. Due to the negative sign of the scalar field perturbations, the linear growth
factor in clustering quintessence models is smaller than that obtained in homogeneous quintessence models (top panel of
figure~\ref{fig:growth_factor}).

As next step we calculated the SCM parameters $\delta_{\rm c}$ and $\Delta_{\rm vir}$ in the context of homogeneous and
clustering non-minimally coupled quintessence models. Due to the negative sign of the scalar field perturbations, we
notice that the linear overdensity $\delta_{\rm c}$ is smaller with respect to the case when the scalar field is
homogeneous (see figure~\ref{fig:delta_c}). The redshift evolution of the virial overdensity $\Delta_{\rm vir}$
parameter shows that in both homogeneous and clustering quintessence models the scalar field sector slows down the
collapse and the formation of overdense regions as it happens for DE models in standard GR gravity (see
figure~\ref{fig:deltavir1}).

Wt this point of the discussion, we want to stress the point that the results here obtained for both the linear growth
factor ($D(a)$) and linear overdensity parameter ($\delta_{\rm c}$) are more general than those in \cite{Pace2014},
where the effective gravitational constant was approximated to $G_{\rm eff}\simeq G_{\rm N}/F$ for a direct comparison
with N-body simulations.

We finally investigated the number count of haloes for non-minimally coupled quintessence models taking into account
the scalar field perturbations. In the case of homogeneous models, we showed that all the models here investigated
give an excess of structures with respect to the concordance $\Lambda$CDM model. The differences between non-minimally
coupled
models and the $\Lambda$CDM model are more pronounced for high masses and high redshifts. We also notice that the
number of haloes in non-minimally coupled quintessence models with negative (positive) coupling parameter $\xi$ is
higher (lower) than that obtained in the minimally coupled quintessence case (see figure~\ref{fig:mass_fun1}).

When the scalar field is clustering, we should in principle modify the mass of the halo. When doing this, we obtain
similar results to the homogeneous case, except for model (1). In model (1), we see that the number of haloes is lower
than in the reference $\Lambda$CDM universe. Since scalar field perturbations are negative (see
figure~\ref{fig:epsilon}), the predicted number of haloes is smaller compared to the homogeneous models (see
tables~\ref{tab:number1} and~\ref{tab:number2}).

\bibliographystyle{mnras}
\bibliography{STscm.bbl}

\appendix

\section{Useful equations in terms of the Planck mass}
Since different works in literature use different definitions of the function F characterising the coupling between the
scalar field and the Ricci scalar, for the sake of completeness, here we write the most important equations used in
this work relaxing the common notation $8\pi G=1$. This will help the reader to implement the following equations
correctly from a dimensional point of view. We also remind the reader that, expressing physical quantities in units of
mass ($[M]$), the density and the scalar field potential have units of $[M^4]$, the Hubble function and time
derivatives have units of $[M]$ and finally the scalar field is expressed in units of the reduced Planck mass
($M_{\rm pl}^2=1/(8\pi G)$), where we have set $\hbar$ (Planck constant) and $c$ (speed of light) to unity. The
function $F$ and the coupling constant $\xi$ are assumed to be dimensionless quantities.

The function $F(\Phi)$ is therefore
\begin{equation}
 F(\phi)=1+\xi\left[\left(\frac{\Phi}{M_{\rm pl}}\right)^2-\left(\frac{\Phi_0}{M_{\rm pl}}\right)^2\right]\;,
\end{equation}
and the density and the pressure of the scalar field are
\begin{eqnarray}
 \rho_{\Phi} & = & \frac{1}{2}\dot{\Phi}^2+U(\Phi)-3HM_{\rm pl}^2\dot{F}\;,\\
 p_{\Phi} & = & \frac{1}{2}\dot{\Phi}^2-U(\Phi)+M_{\rm pl}^2(\ddot{F}+2H\dot{F})\;.
\end{eqnarray}
Hence Friedmann equations read
\begin{eqnarray}
 3F(\Phi) H^2  & = & 8\pi G\left(\rho_{\rm m}+\frac{1}{2}\dot{\Phi}^2+U\right)-3H\dot{F}\;,\label{eqn:H2}\\
 -2F(\Phi) \dot H & = & 8\pi G\left(\rho_{\rm m}+\dot{\Phi}^2\right)+\ddot{F}-H\dot{F}\;,\label{eqn:Hdot}\\
 \frac{\ddot{a}}{a} & = & -\frac{4\pi G}{3F}\left[\rho_{\rm m}+2\dot{\Phi}^2-2U\right.\nonumber\\
                    & &   +\left.3M_{\rm pl}^2(\ddot{F}+H\dot{F})\right]\;.\label{eqn:addot}
\end{eqnarray}
Note that equation~(\ref{eqn:addot}) corrects a typo in equation~22 of \cite{Pace2014}.

The Klein-Gordon equation can be written in a more general form by noticing that $R=6(2H^2+\dot{H})$:
\begin{equation}
 \ddot{\Phi}+3H\dot{\Phi}+\frac{dU}{d\Phi}=\frac{1}{2}M_{\rm pl}^2\frac{dF}{d\Phi}R\;.
\end{equation}

The two different forms of the effective gravitational constant read
\begin{eqnarray}
 G_{\rm eff}^{(h)} & = & \frac{G_{\rm N}}{F}
                         \left(\frac{2F+2M_{\rm pl}^2F_{,\Phi}^2}{2F+3M_{\rm pl}^2F_{,\Phi}^2}\right)
                     =   \frac{G_{\rm N}}{F}\frac{2+2\omega_{\rm JDB}^{-1}}{2+3\omega_{\rm JDB}^{-1}}\;,\\
 G_{\rm eff}^{(p)} & = & \frac{G_{\rm N}}{F}
                         \left(\frac{2F+4M_{\rm pl}^2F_{,\Phi}^2}{2F+3M_{\rm pl}^2F_{,\Phi}^2}\right)
                     =   \frac{G_{\rm N}}{F}\frac{2+4\omega_{\rm JDB}^{-1}}{2+3\omega_{\rm JDB}^{-1}}\;,
\end{eqnarray}
where we introduced the {\it Jordan-Brans-Dicke} parameter $\omega_{\rm JDB}$:
\begin{equation}\label{eqn:JBD}
 \omega_{\rm JDB}^{-1}=M_{\rm pl}^2F_{,\Phi}^2/F\;.
\end{equation}
By using equation~(\ref{eqn:JBD}), we see that GR is recovered for $\omega_{\rm JDB}\gg 1$. In addition, we also
notice that for $\omega_{\rm JDB}\gg 1$ we obtain $G_{\rm eff}^{(h)}=G_{\rm eff}^{(p)}\simeq G_{\rm N}/F$. This shows
once again, from a different point of view, that in the general relativistic regime, fluctuations of the scalar field
become unimportant.

\label{lastpage}

\end{document}